\documentclass[aps,prd,preprintnumbers,showkeys,nofootinbib,
superscriptaddress,fleqn,floatfix,tightenlines,10pt]{revtex4-2}
\usepackage{amsmath,amsfonts,amssymb,amscd,amsxtra,amsthm}
\usepackage{graphicx}  
\usepackage{epstopdf}
\usepackage{dcolumn}  
\usepackage{bm}          
\usepackage{slashed}
\usepackage{float}
\usepackage{cancel} 
\usepackage{mathtools}
\usepackage{amsbsy}
\usepackage{amstext}
\usepackage{threeparttable}

\usepackage[utf8]{inputenc}  
\usepackage{booktabs}
\usepackage[normalem]{ulem} 
\usepackage[dvipsnames,table,xcdraw]{xcolor} 
\usepackage{tabularx}
\usepackage{enumitem}   
\usepackage{array} 
\usepackage{tikz}
\usepackage{multirow}
\usepackage[symbol]{footmisc}

\usepackage{rotating}
\renewcommand\sout{\bgroup \color{red} \ULdepth=-.5ex \ULset}

\makeatletter
 
\begin{document}  
\preprint{INHA-NTG-04/2022}
\title{Axial-vector transition form factors of the singly heavy
  baryons} 
\author{Jung-Min Suh}
\email[E-mail: ]{suhjungmin@inha.edu}
\affiliation{Department of Physics, Inha University, Incheon 22212,
Republic of Korea}

\author{Hyun-Chul Kim}
\email[E-mail: ]{hchkim@inha.ac.kr}
\affiliation{Department of Physics, Inha University, Incheon 22212,
Republic of Korea}
\affiliation{School of Physics, Korea Institute for Advanced Study
(KIAS), Seoul 02455, Republic of Korea}
\date{\today}
\begin{abstract} 
We investigate the axial-vector transition form factors of the
lowest-lying singly heavy baryons within the framework of the chiral
quark-soliton model. We consider the linear $m_{\mathrm{s}}$ corrections,
dealing with the strange current quark mass $m_{\mathrm{s}}$ as a
small perturbation. Since we have various relations between different
transitions because of isospin symmetry and flavor SU(3) symmetry
breaking, only two axial-vector transition form factors are
independent. We present the numerical results for these form factors. 
The effects of the flavor SU(3) symmetry breaking turn out tiny, so we
neglect them. We also compute the decay rates for several strong
decays of the singly heavy baryons and compare the results with the
experimental data and those from other models. While the results for the
$\Sigma_c\to \Lambda_c^++\pi$ and $\Sigma_c^*\to \Lambda_c^++\pi$
decays are slightly overestimated in comparison with the corresponding
experimental data, those for the $\Xi_c^*\to \Xi_c+\pi$ are in
remarkable agreement with the data. 
\end{abstract}
\maketitle

\section{Introduction}
The structure of singly heavy baryons has been much less known than that of
the light baryons, both experimentally and theoretically. Even for the
charmed baryons in the ground states, we know only their masses and
decay widths~\cite{PDG}. Recently, the 
electromagnetic properties of the singly heavy baryons have been
investigated within lattice QCD~\cite{Can:2013tna, Bahtiyar:2015sga, 
  Bahtiyar:2016dom, Bahtiyar:2019ykq}. There have also been various
theoretical works on their electromagnetic structure. On the other
hand, there are very few works on the axial-vector properties of the
singly heavy baryons. Since the LHCb Collaborations have continuously
announced a series of new experimental data on
the heavy baryons~\cite{Aaij:2012da, Aaij:2013qja,
  Aaij:2014esa, Aaij:2014lxa, Aaij:2014yka, Aaij:2017nav,
  LHCb:2020gge, LHCb:2021ptx}, one may expect that future experiments
will will reveal the axial-vector structure of the heavy baryons.  
 
\begin{figure}[ht]
  \includegraphics[scale=0.9]{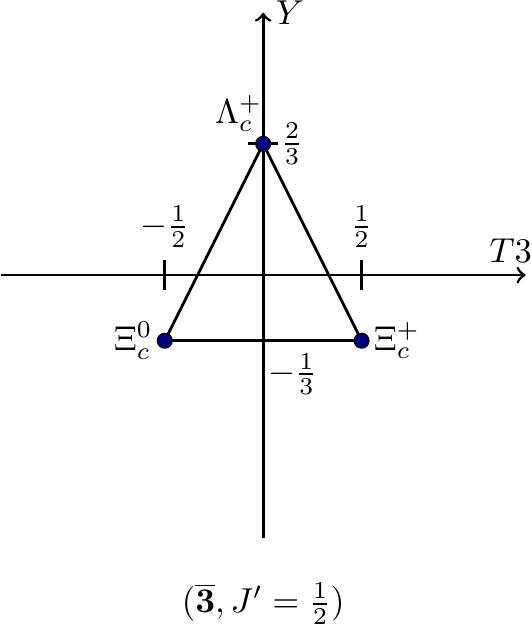}
  \includegraphics[scale=0.9]{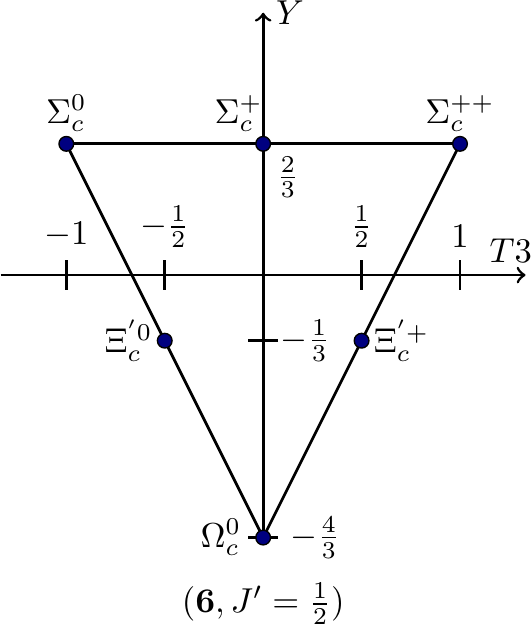} 
  \includegraphics[scale=0.9]{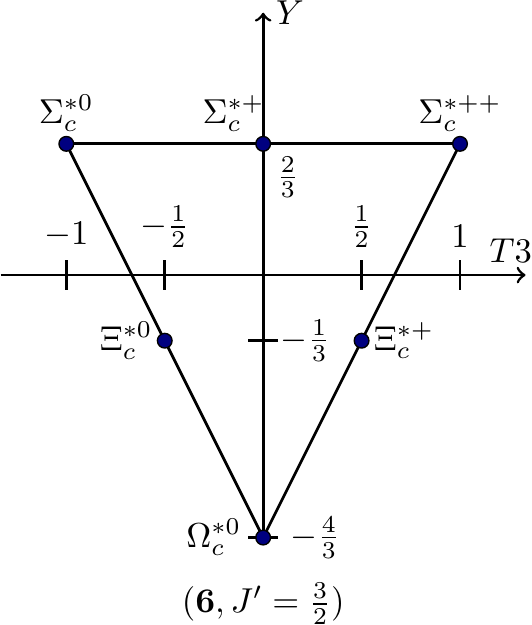}
\caption{Flavor $\mathrm{SU(3)_f}$ representation of the singly heavy
  charmed baryons}  
\label{fig:1}
\end{figure}

A singly heavy baryon consists of a heavy quark and two light
quarks.  In the limit of the infinitely heavy-quark mass ($m_Q\to
\infty$), the spin of the heavy quark $\bm  S_Q$ is conserved, 
which brings about the conservation of the spin of the
light-quark degrees of freedom: $\bm{S}_{\mathrm{L}} \equiv
\bm{S}-\bm{S}_Q$~\cite{Isgur:1989vq, Isgur:1991wq,
  Georgi:1990um}. It is called the heavy-quark spin
symmetry, which allows one to take the total spin of the light 
quarks as a good quantum number. Thus, we can classify the singly
heavy charmed baryons in the ground states according to the
representation of flavor $SU(3)_{\mathrm{f}}$ symmetry:  
$\bm{3}\otimes \bm{3}=\overline{\bm{3}} \oplus \bm{6}$, where
the baryon anti-triplet ($\overline{\bm{3}}$) has $S_{\mathrm{L}}=0$
and $S=1/2$, whereas the baryon sextet ($\bm{6}$) carries  
$S_{\mathrm{L}}=1$. Since the spin of the heavy quark is coupled to
$\bm{S}_L$, the baryon sextet have two degenerate representations with
$S=1/2$ and $S=3/2$, respectively, as illustrated in 
Fig.~\ref{fig:1}. This degeneracy is removed by introducing the color
hyperfine interaction in order $1/m_Q$. 

In the limit $m_Q\to \infty$, we regard the heavy quark inside a
heavy baryon as a static color source, so the light quarks 
govern the structure of the singly heavy baryons. 
Some years ago, Yang et al.~\cite{Yang:2016qdz} proposed a pion
mean-field approach to explain the masses of singly heavy baryons,
following the idea proposed by Ref.~\cite{Diakonov:2010tf}.  
Witten showed in his seminal 
papers~~\cite{Witten:1979kh,Witten:1983tx} that in the limit of the    
large number of colors ($N_c\to\infty$) a baryon arises as a bound
state of $N_c$ \textit{valence} quarks in a pion mean field with 
a hedgehog symmetry~\cite{Pauli:1942kwa, Skyrme:1961vq} that is a
minimal extension of spherical symmetry with the characteristics of
the pions considered. Since the quantum fluctuation around the saddle
point of the pion field is suppressed by $1/N_c$ factor, one can
ignore it. In this large $N_c$ limit, the presence of $N_c$   
valence quarks that constitute the lowest-lying baryons
causes the vacuum polarization, which creates the pion mean
field. This pion mean field makes \emph{self-consistently} the
$N_c$ valence quarks bound. To keep the hedgehog symmetry preserved in
the case of flavor $\mathrm{SU(3)}_{\mathrm{f}}$, an $\mathrm{SU(2)}$
soliton is embedded into the isospin corner of
$\mathrm{SU(3)}_{\mathrm{f}}$~\cite{Witten:1983tx}. 

The chiral quark-soliton model ($\chi$QSM)~\cite{Diakonov:1987ty,
 Wakamatsu:1990ud, Diakonov:1997sj} was constructed such that
it realizes Witten's idea. Note that in the $\chi$QSM the right
hypercharge $Y_R=N_c/3$ is constrained by the $N_c$
valence quarks, which is distinguished from the Skyrme model where
the Wess-Zumino-Witten term fixes it. This indicates that the explicit
valence quark degrees of freedom determine properly the baryon
representations in the $\chi$QSM: This constrained right hypercharge
selects allowed representations of light baryons such as the baryon
octet ($\bm{8}$), the decuplet ($\bm{10}$), etc. The $\chi$QSM has 
great merit because it can extend directly to describe singly heavy 
baryons. In the limit of $m_Q\to\infty$, a heavy quark inside the
singly heavy baryon stays dormant but can play a role only as a static 
color source so that a colored soliton consisting of $N_c-1$ valence
quarks emerges. The right hypercharge $Y_R=(N_c-1)/3$ is constrained
by the $N_c-1$ valence quarks, which picks the allowed representations
of singly heavy baryons such as the baryon antitrplet 
($\overline{\bm{3}}$) and the baryon sextet ($\bm{6}$) as depicted in
Fig.~\ref{fig:1}, and in addition the baryon
antidecapentaplet ($\overline{\bm{15}}$)~\cite{Kim:2017jpx,  
  Kim:2017khv}. This extended $\chi$QSM have successfully applied in
describing properties of the singly heavy baryons such as the mass
splittings~\cite{Yang:2016qdz, Kim:2018xlc, Kim:2019rcx}, 
isospin mass differences~\cite{Yang:2020klp}, 
magnetic moments~\cite{Yang:2018uoj},  
magnetic transitions and radiative decays~\cite{Yang:2019tst}, 
electromagnetic and radiative transition form
factors~\cite{Kim:2018nqf, Kim:2020uqo, Kim:2019wbg, Kim:2021xpp}, 
and gravitational form factors~\cite{Kim:2020nug}.

In the present work, we investigate the axial-vector
transition form factors of the low-lying singly charmed baryons,
including both the strangeness-conserving ($\Delta 
S=0)$ and strangeness-changing ($|\Delta S|=1$) transitions. 
While there have been no theoretical works on the axial-vector
transition form factors of singly heavy baryons, 
many theoretical groups studied their decay widths:
for example, heavy hadron chiral perturbation theory (HH$\chi$PT) 
~\cite{Yan:1992gz,Huang:1995ke,Cheng:2015naa,Pirjol:1997nh}, a quark
model (QM)~\cite{Rosner:1995yu}, the light-front quark
model (LFQM)~\cite{Tawfiq:1998nk}, the relativistic three-quark
model (RTQM)~\cite{Ivanov:1999bk}, the nonrelativistic constituent quark 
models (NRQM)~\cite{Albertus:2005zy,Nagahiro:2016nsx}, 
the ${}^{3} P_{0}$ strong decay model (${}^{3}
P_{0}$)~\cite{Chen:2007xf}, light cone QCD sum rules
(LQCDSR)~\cite{Azizi:2008ui} and lattice QCD(LQCD)~\cite{Can:2016ksz}. 
Since the heavy quark is not involved in the present axial-vector
transitions of the singly heavy quarks, we can concentrate on the
light quark degrees of freedom to compute the axial-vector transition
form factors of the singly charmed baryons. We consider the rotational
$1/N_c$ corrections and explicit breaking of
$\mathrm{SU(3)}_{\mathrm{f}}$ symmetry to linear
order~\cite{Christov:1995vm}. Since we have already computed 
the axial-vector transition form factors of the baryon 
decuplet~\cite{Suh:2022guw}, we will focus on how the axial-vector
transition form factors of the singly charmed baryons with spin 3/2
behave differently from those of the $\Delta$ isobar. 

The structure of the current work is summarized as follows:
In section II, we define the axial-vector transition form factors from
the baryon sextet to both the baryon antitriplet and sextet, based on
the transition matrix elements of the axial-vector current.
In Section III, we show how to compute the axial-vector transition
form factors of the singly heavy baryons in the $\chi$QSM. In Section
IV, we first compare the present results with that of the $\Delta\to
p$ axial-vector transition form factor. We then discuss the effects of 
$\mathrm{SU(3)}_{\mathrm{f}}$ symmetry breaking. The last section is
devoted to summary and conclusions of the present work. 

\section{Axial-vector transition form factors between the singly 
heavy baryons} 
\label{sec:2}
The axial-vector current of a singly heavy baryon consists of the
light-quark and heavy-quark parts: 
\begin{align}
A^{\chi}_\mu (x) = \bar{\psi} (x) \gamma_\mu \gamma_{5} 
\frac{\lambda^{\chi}}{2} \psi(x) +\bar{\Psi}(x) \gamma_{\mu}\gamma_{5}
  \Psi(x), 
\label{eq:AxialCUR}
\end{align}
where $\psi(x)$ represents the light-quark field $\psi=(u,d,s)$ in
flavor space and $\Psi(x)$ denotes the heavy-quark field generically
for the charm or bottom quark. The $\lambda^{{\chi}}$ denotes the
well-known $\mathrm{SU(3)}_{\mathrm{f}}$ Gell-Mann matrices for which
the index $\chi$ is determined by strangeness-conserving $\Delta S=0$
transitions ($\chi=1\pm i2$) and for $|\Delta S|=1$ ones ($\chi=4 \pm
i5$), respectively. Considering the Lorentz structure together with
spin, parity, time reversal, and charge conjugation, we can parametrize
the transition matrix elements of the axial-vector current 
between the baryons with spin 1/2 in terms of two different real form
factors:  
\begin{align}
\langle B'_{\frac{1}{2}}(p',J_3') | A_\mu^{\chi}(0) |
  B_{\frac{1}{2}}(p,J_3) \rangle  
&= \overline{u}(p',J_3') \left[ G_{A}^{(\chi)}(q^2) \gamma_{\mu}
  +\frac{G_{P}^{(\chi)}(q^2)}{M_{B'}+M_{B}} q_{\mu} \right]
  \frac{\gamma_{5}}{2} u(p,J_3),  
  \label{eq:MatrixEl1}
\end{align}
where $G_A^{(\chi)}$ and $G_P^{(\chi)}$ are the axial-vector
transition and pesudoscalar transition form factors of the
corresponding baryon sextet with spin 1/2, respectively. 
$u(p,J_3)$ and $\overline{u}(p',J_3')$ stand for the Dirac spinors for 
the initial and final baryon states, respectively. $M_{B}$ and $M_{B'}$
designate the corresponding masses, respectively. $q_\mu$ denotes the
momentum transfer and $q^2$ its square. 
The transition matrix elements between the baryon with spin 3/2 and
with spin 1/2 are parametrized in terms of four real form
factors~\cite{Adler:1968tw}:   
\begin{align}
\langle B'_{\frac{1}{2}}(p',J_3') | A_\mu^{\chi}(0) |
  B_{\frac{3}{2}}(p,J_3) \rangle  
&=  \overline{u}(p',J_3') \left[ \left 
  \{\frac{C_{3}^{A(\chi)}(q^2)}{M_{B'}} \gamma^{\nu} 
    + \frac{C_{4}^{A(\chi)}(q^2)}{M_{B'}^{2}} p^{\nu}\right \}
  (g_{\alpha \mu}g_{\rho \nu}-g_{\alpha \rho}g_{\mu \nu})q^{\rho} \right. \cr
&\hspace{2cm} \left. + \,C_{5}^{A(\chi)}(q^2)g_{\alpha \mu} 
  +\frac{C_{6}^{A(\chi)}(q^2)}{M_{B'}^2}q_{\alpha}q_{\mu}
  \right ] u^{\alpha}(p,J_3), 
  \label{eq:MatrixEl2}
\end{align}
where $g_{\alpha \beta}$ represents the metric tensor 
$g_{\alpha \beta} =\mathrm{diag}(1,\,-1,\,-1,\,-1)$.
In the rest frame of a initial baryon, $p^{\alpha}$,
$u^\alpha (p,\,J_3)$ is the Rarita-Schwinger spinor that describes 
a baryon with spin 3/2, carrying the momentum $p$ and $J_3$, which
can be described by the combination of the polarization vector and the
Dirac spinor, $u^\alpha (p,\,J_3)= \sum_{i,s} C^{\frac{3}{2} J_3}_{1i\,\frac{1}{2}s} 
\epsilon^{\alpha}_{i}(p) u_{s}(p)$. It satisfies the Dirac equation
and the auxiliary equations $p_{\alpha} u^\alpha(p,J_3)=0$ and $\gamma_\alpha
u^\alpha(p,J_3)=0$~\cite{Rarita:1941mf}. 
The momenta of the initial and final states $p$ and $p'$, and the
momentum transfer are explicitly written as 
\begin{align}
&p=(M,\bm{0}),\quad p'=(E',-\bm{q}), \quad q=(\omega_{q},\bm{q})
  \label{eq:Momenta}
\end{align}
where $q^2=-Q^2$ with $Q^{2}>0$. 
Thus, the three-vector momentum and energy of the momentum transfer
are expressed by 
\begin{align}
&|\vec{q}|^{2}=\left(\frac{M_{B}^{2}+M_{B'}^{2}+Q^{2}}{2M_{B}}\right)^{2}-M_{B'}^{2}, 
  \quad 
  \omega_{q}=\left(\frac{M_{B'}^{2}-M_{B}^{2}+Q^{2}}{2M_{B}} \right). 
  \label{eq:VecQ}
\end{align}

The axial-vector transition form factor $G_{A,\, B \rightarrow
  B'}^{(\chi)}(Q^2)$ can be obtained in terms of the spatial parts of
the axial-vector current in the spherical tensor form
\begin{align}
G_{A,\, B \rightarrow B'}^{(\chi)}(Q^{2}) =&
  -2\sqrt{\frac{M_{B'}}{E_{B'}+M_{B'}}} \; 
  \left[\int d^{3}r j_{0}(|\bm{q}||\bm{r}|)\langle
   B'_{\frac{1}{2}}(p',S_{3}^{'})|A_{10}^\chi(\bm{r}) 
  |B_{\frac{1}{2}}(p,S_{3}) \rangle \right.\cr
  &\left.-\int d^{3}r j_{2}(|\bm{q}||\bm{r}|)\langle
    B'_{\frac{1}{2}}(p',S_{3}^{'})|\{Y_{2} \otimes
    A_{1}^\chi(\bm{r})\}_{10} 
  |B_{\frac{1}{2}}(p,S_{3}) \rangle \right].
  \label{eq:GA}
\end{align}
Since the form factor $C_{5,\, B \rightarrow B'}^{A(\chi)}(q^{2})$ is
the most dominant one, we concentrate on it. Its expression is very
similar to 
Eq.~\eqref{eq:GA} 
\begin{align}
C_{5,\, B \rightarrow B'}^{A(\chi)}(Q^{2}) =&
    -\sqrt{\frac{2M_{B'}}{E_{B'}+M_{B'}}} \; 
  \left[\int d^{3}r j_{0}(|\bm{q}||\bm{r}|)\langle
  B'_{\frac{1}{2}}(p',S_{3}^{'})|   A_{10}^\chi(\bm{r}) 
  |B_{\frac{3}{2}}(p,S_{3}) \rangle \right.\cr
  &\left.-\int d^{3}r j_{2}(|\bm{q}||\bm{r}|)\langle
    B'_{\frac{1}{2}}(p',S_{3}^{'})|\{Y_{2} \otimes
    A_{1}^\chi(\bm{r})\}_{10} 
  |B_{\frac{3}{2}}(p,S_{3}) \rangle\right]  .
  \label{eq:C5}
\end{align}
Note that $G_{A,\, B\to B'}^{(\chi)}(0)$ and $C_{5,\, B \rightarrow B'}^{A(\chi)}(0)$
are related to the strong coupling constants $g_{\pi BB'}$ by the
Goldberger-Treiman relations. 

\section{A singly heavy baryon in the chiral quark-soliton model}
\label{sec:3}
The $\chi$QSM has proved great merit by showing that it can describe
both the light and singly heavy baryons on the same footing. 
Since we want to discuss the axial-vector transition form factors of
the singly heavy baryons in this work, we will first 
explain how a singly heavy baryon can be formulated in the pion
mean-field approach. Let us define the normalization of the baryon
state as $\langle B(p',J_3') | B (p,J_3)\rangle = 2 p_0
\delta_{J_3'J_3} (2\pi)^{3}\delta^{(3)}(\bm{p}'-\bm{p})$. In the large
$N_c$ limit, this normalization is reduced to $\langle
B(p',J_3') | B (p,J_3)\rangle = 2 M_B \delta_{J_3'J_3}
(2\pi)^{3}\delta^{(3)}(\bm{p}'-\bm{p})$, 
where $M_B$ is a baryon mass. 
A singly heavy baryon consists of the $N_c-1$ valence quarks and
a static heavy quark, so the corresponding state can be written in
terms of the Ioffe-type current of the $N_c-1$ valence quarks and a
heavy-quark field in Euclidean space as follows: 
\begin{align}
|B,p\rangle &= \lim_{x_4\to-\infty} \exp(ip_{4}x_{4})
              \mathcal{N}(\bm{p}) \int d^3 x
              \exp(i\bm{p}\cdot \bm{x}) (-i\Psi_h^\dagger(\bm{x}, x_4)
              \gamma_4) J_B^\dagger (\bm{x},x_4) 
              |0\rangle,\cr
\langle B,p| &= \lim_{y_4\to \infty} \exp(-ip'_4 y_4) 
               \mathcal{N}^*(\bm{p}') \int d^3 y
              \exp(-i\bm{p}'\cdot \bm{y}) \langle 0| J_{B}
                 (\bm{y},y_4) \Psi_h (\bm{y},y_4), 
\end{align}
where $\mathcal{N}(\bm{p}) (\mathcal{N}^*(\bm{p}'))$ represents 
the normalization factor depending on the initial (final) momentum. 
$J_B(x)$ and $J_B^\dagger(y)$ stand for the Ioffe-type current
consisting of the $N_c-1$ valence quarks~\cite{Diakonov:1987ty}
defined by  
\begin{align}
J_B(x) &= \frac1{(N_c-1)!} \epsilon_{\alpha_1\cdots \alpha_{N_c-1}} 
\Gamma_{(TT_3Y)(JJ_3Y_R)}^{f_1\cdots  f_{N_c-1}}
\psi_{f_1 \alpha_1}(x)\cdots \psi_{f_{N_c-1} \alpha_{N_c-1}}(x),\cr  
J_{B}^\dagger(y) &= \frac1{(N_c-1)!} \epsilon_{\alpha_1\cdots \alpha_{N_c-1}}
\Gamma_{(TT_3Y)(JJ_3'Y_R)}^{f_1\cdots f_{N_c-1}}
(-i\psi^\dagger(y)\gamma_4)_{f_1\alpha_1} \cdots
  (-i\psi^\dagger(y)\gamma_4)_{f_{N_c-1}\alpha_{N_c-1}} ,
\end{align}
where $f_1\cdots f_{N_c-1}$ and $\alpha_1\cdots\alpha_{N_c-1}$ denote
respectively the spin-isospin and color
indices. $\Gamma_{(TT_3Y)(JJ_3Y_R)}$ correspond to matrices with the
quantum numbers $(TT_3Y)(JJ_3Y_{R})$ for a given state. 
In the $\chi$QSM, right hypercharge $Y_R$ is constrained by the number
of the valence quarks while it is fixed by the Wess-Zumino-Witten
term in the Skyrme model. It provides a distinct advantage for the
$\chi$QSM, since the right hypercharge $Y_{R}$ for singly heavy baryons 
is determined by the $N_c-1$ valence quarks: $Y_{R}=(N_c-1)/3$. The
right hypercharge $Y_{R}=2/3$ with $N_c=3$ grants the 
baryon antitriplet ($\overline{\bm{3}}$), sextet ($\bm{6}$),
antipentadecaplet ($\overline{\bm{15}}$), and so
on~\cite{Yang:2016qdz, Kim:2017jpx, Kim:2017khv,
  Kim:2018cxv}. $\psi_{f_k \alpha_k}(x)$ is the light-quark
field. $\Psi_h(x)$ denotes the heavy-quark field, making 
the singly heavy baryon a color singlet. In the limit of $m_Q\to
\infty$, the singly heavy baryon complies with the heavy-quark flavor
symmetry, so that the heavy-quark field can be written as 
\begin{align}
\Psi_h(x) = \exp(-im_Q v\cdot  x) \tilde{\Psi}_h(x). 
\end{align}
Here $\tilde{\Psi}_h(x)$ stands for a rescaled heavy-quark field almost on
mass-shell. It carries no information on the heavy-quark mass in the
leading-order approximation in the heavy-quark expansion. $v$ is 
the velocity of the heavy quark~\cite{Isgur:1989vq, Isgur:1991wq}
with the superselection rule~\cite{Georgi:1990um}.  

We now show the normalization factor
$\mathcal{N}^*(\bm{p}')\mathcal{N}(\bm{p})$ to be $2 M_B$. The
normalization of the baryon state can be computed as follows:       
\begin{align}
 \label{eq:normal}
\langle B(p',J_3') | B (p,J_3)\rangle &=
\frac{1}{\mathcal{Z}_{\mathrm{eff}}}
  \mathcal{N}^*(p')\mathcal{N}(p)
\lim_{x_4\to -\infty}  \lim_{y_4\to \infty}   
\exp\left(-iy_4p_4'+ix_4p_4\right)  \cr
& \times \int  d^3x d^3y 
  \exp(-i\bm{p}'\cdot \bm{y}+  i\bm{p}\cdot \bm{x})
 \int \mathcal{D} U  \mathcal{D} \psi
  \mathcal{D}\psi^\dagger \mathcal{D} \tilde{\Psi}_h \mathcal{D}
  \tilde{\Psi}_h^\dagger \cr  
&\times J_B (y)\Psi_h(y) (-i\Psi_h^\dagger(x) \gamma_4) J_B^\dagger(x) \cr
&\times \exp\left[\int d^4z\left\{
  (\psi^\dagger(z))_{\alpha}^{f} \left( i\rlap{/}{\partial}  + i
  MU^{\gamma_5} + i  \hat{m} \right)_{fg} \psi^{g\alpha}(z) +
  \Psi_h^\dagger(z) v\cdot \partial 
  \Psi_h(z) \right\} \right] \cr
&=
  \frac{1}{\mathcal{Z}_{\mathrm{eff}}}\mathcal{N}^*(p')\mathcal{N}(p)
  \lim_{x_4\to -\infty} \lim_{y_4\to \infty} 
 \exp\left(-iy_4p_4'+ix_4p_4\right) \cr
& \times \int  d^3x d^3y 
  \exp(-i\bm{p}'\cdot y+  i\bm{p}\cdot \bm{x}) \langle 
J_B (y) \Psi_h(y) (-i\Psi_h^\dagger(x) \gamma_4) J_B^\dagger(x)\rangle_0, \cr
\end{align}
where $\mathcal{Z}_{\mathrm{eff}}$ represents the low-energy effective
QCD partition function with the quark fields integrated out
\begin{align}
\mathcal{Z}_{\mathrm{eff}} = \int   \mathcal{D} U \exp(-S_{\mathrm{eff}}).    
\end{align}
$\langle ... \rangle_{0}$ in Eq.~\eqref{eq:normal} designates the
vacuum expectation value of the baryon correlation function. 
$S_{\mathrm{eff}}$ is known as the effective chiral action 
(E$\chi$A) defined by  
\begin{align}
S_{\mathrm{eff}} = -N_c \mathrm{Tr} \ln \left[ i\rlap{/}{\partial}  +
  iMU^{\gamma_5} + i   \hat{m}\right], 
\end{align} 
which embraces the effective nonlocal interaction between the quark
and pseudo-Nambu-Goldstone (pNG) fields. $M$ is the dynamical quark mass that
arises from the spontaneous breakdown of chiral symmetry.
The $U^{\gamma_5}$ stands for the chiral field that is defined by
\begin{align}
  \label{eq:1}
U^{\gamma_5} (z) = \frac{1-\gamma_5}{2}   U(z) + U^\dagger(z)
  \frac{1+\gamma_5}{2} 
\end{align}
with
\begin{align}
  \label{eq:2}
U(z) = \exp[{i\pi^a(z) \lambda^a}],  
\end{align}
where $\pi^a(z)$ represents the pNG fields and $\lambda^a$ are the
flavor Gall-Mann matrices. $\hat{m}$ displays the mass
matrix of current quarks $\hat{m} = \mathrm{diag}(m_{\mathrm{u}},\,
m_{\mathrm{d}},\,m_{\mathrm{s}})$.
We regard the strange current quark mass $m_{\mathrm{s}}$ as a small 
perturbation. 

The Green function of a light quark in the
$\chi$QSM~\cite{Diakonov:1987ty} is given by    
\begin{align}
G(y,x) &= \left \langle y \left |
                      \frac1{i\rlap{/}{\partial} + i 
  MU^{\gamma_5}  +i\overline{m}} (i\gamma_4)\right | x\right 
\rangle \cr
&= \Theta(y_4   -x_4) 
\sum_{E_n>0} e^{-E_n(y_4-x_4)} \psi_n(\bm{y})
  \psi_n^\dagger(\bm{x}) \cr
  &- \Theta(x_4  -y_4) 
\sum_{E_n<0} e^{-E_n(y_4-x_4)}
  \psi_n(\bm{y})\psi_n^{\dagger}(\bm{x}), 
\end{align}
where $\Theta(y_4   -x_4)$ is the Heaviside step function. 
$\overline{m}$ denotes the average mass of the 
up and down current quarks: $\overline{m}=(m_{\mathrm{u}}+
m_{\mathrm{d}})/2$ that constitutes an essential part in producing
the correct Yukawa tail of the soliton profile function.  
$E_n$ corresponds to the energy eigenvalue of the single-quark
eigenstate given by
\begin{align}
H \psi_n(\bm{x}) = E_n \psi_n(\bm{x}),  
\end{align}
where $H$ is the one-body Dirac Hamiltonian in the presence of
the pNG boson fields, defined by
\begin{align}
  H = \gamma_4 \gamma_i \partial_i + \gamma_4 MU^{\gamma^5}
    + \gamma_4 \bar{m} \mathbf{1}.
\end{align}
The Green function for the heavy quark in the limit of $m_Q\to\infty$
is given by the Heaviside step function and Dirac delta function 
\begin{align}
G_h (y,x) = \left \langle y \left |   \frac1{\partial_4 }\right |
  x\right \rangle= \Theta(y_4-x_4) \delta^{(3)}(\bm{y} - \bm{x}) ,  
\end{align}
which is the natural form of the heavy-quark propagator in the
$m_Q\to\infty$ limit. 
Using these Green functions for the light and heavy quarks and taking
the limit of $y_4-x_4=T\to\infty$, we arrive at expression for the
baryon correlation function $\langle J_B (y)\Psi_h(y)
(-i\Psi^{\dagger}_h(x)\gamma_4)
J_B^\dagger(x)\rangle_0$~\cite{Kim:2021xpp}:
\begin{align}
\langle J_B (y)\Psi_h(y) (-i\Psi_h^\dagger(x) \gamma_4)
J_B^\dagger(x)\rangle_0 &  \sim \exp\left[-\{(N_c-1)E_{\mathrm{val}} +
  E_{\mathrm{sea}}+m_Q\}T\right] = \exp[-M_B T].  
\label{eq:baryon_corr}
\end{align}
Since the result for the correlation function given in
Eq.~\eqref{eq:baryon_corr} is canceled with the term
$\exp\left(-iy_4p_4'+ix_4p_4\right)= \exp[M_BT] $ in the large $N_c$ 
limit, i.e., $-ip'_{4}=-ip_{4}=M_{B}=\mathcal{O}(N_{c})$. Thus, the
normalization factor is reduced to the mass of a singly heavy
baryon: $\mathcal{N}^*(\bm{p}')\mathcal{N}(\bm{p})=2M_{B}$.  
Combining Eq.~\eqref{eq:baryon_corr} with the normalization constant,
we find that the classical mass of the singly heavy baryon is 
given by the sum of the $N_c-1$ soliton and the heavy-quark masses 
\begin{align}
M_B = (N_c-1) E_{\mathrm{val}} + E_{\mathrm{sea}} + m_Q,  
\label{eq:bmass}
\end{align}
which was assumed in a previous work~\cite{Kim:2018xlc}. The classical
mass given in Eq.~\eqref{eq:bmass} comes into critical play, when we
derive the axial-vector transition form factors of the singly heavy
baryons, which will be mentioned in Section~\ref{sec:5}.  

\section{Axial-vector transition form factors in the chiral
  quark-soliton model} 
\label{sec:4}
We now show how to compute the transition matrix elements of the
axial-vector current~\eqref{eq:MatrixEl1}, using the functional
integral. Since the heavy quark is not involved, we will only consider
the light quark degrees of freedom 
\begin{align}
&\langle B(p',\, J_3') | A^{a}_\mu(0) |B(p,\,J_3)\rangle =
  \frac1{\mathcal{Z}} \lim_{T\to\infty} \exp\left(i p_4\frac{T}{2} 
  - i p_4' \frac{T}{2}\right) \int d^3x d^3y \exp(-i \bm{p}'\cdot 
  \bm{y} + i \bm{p}\cdot \bm{x})  \cr
  &\int \mathcal{D} \pi^a \int
  \mathcal{D}    \psi \int \mathcal{D} \psi^\dagger 
J_{B}(\bm{y},\,T/2) \psi^\dagger(0) 
  \gamma_4\gamma_\mu \gamma_{5} \frac{\lambda^{a}}{2} \psi(0) 
  J_B^\dagger (\bm{x},\,-T/2) \cr
  &\times \exp\left[-\int d^4 z (\psi^\dagger
  (z))_\alpha^f (i\slashed{\partial} + i MU^{\gamma_5} + i
  \hat{m})_{fg}\psi^{g\alpha} (z)    \psi\right].  
\label{eq:correlftn}
\end{align}
In the large-$N_{c}$ limit, we use the saddle-point approximation
to get the classical soliton. However, we have to take into
account the zero modes that do not change the energy of the
soliton. We assume that the soliton rotates slowly so we deal with
the angular velocities as a perturbative parameter. The integral over
the translational zero modes in the leading order provides naturally
the Fourier transform, which means that the baryon state has the
proper translational symmetry. Thus, the functional integral over the
pNG field is reduced to ordinary integrals over the zero modes. 
We also regard the strange current quark mass $m_{\mathrm{s}}$ as a
perturbation. Having performed the zero-mode quantization, we obtain
the collective Hamiltonian as follows: 
\begin{align}
H_{\mathrm{coll}} = H_{\mathrm{sym}} + H_{\mathrm{sb}},
\end{align}
where
\begin{align}
  \label{eq:Hamiltonian}
H_{\mathrm{sym}} &= M_{\mathrm{cl}} + \frac1{2I_1} \sum_{i=1}^3
  \hat{J}_i^2 + \frac1{2I_2} \sum_{p=4}^7 \hat{J}_p^2,\cr
H_{\mathrm{sb}} &= \alpha D_{88}^{(8)} + \beta \hat{Y} +
  \frac{\gamma}{\sqrt{3}} \sum_{i=1}^3 D_{8i}^{(8)} \hat{J}_i,
\end{align}
where $I_1$ and $I_2$ are the moments of inertia for the classical
soliton and $D^{(8)}_{ab}$ is the SU(3) Wigner $D$ function. 
The inertial parameters $\alpha$, $\beta$ and $\gamma$, which break
flavor SU(3) symmetry explicitly, are written in terms of the
moments of inertia $I_1$ and $I_2$, and the anomalous moments of
inertia $K_1$ and $K_2$
\begin{align}
\alpha=\left (-\frac{\bar{\Sigma}_{\pi N}}{3\overline{m}}
  +\frac{K_{2}}{I_{2}} \bar{Y}\right )m_{\mathrm{s}},
  \quad \beta=-\frac{ K_{2}}{I_{2}}m_{\mathrm{s}}, 
  \quad \gamma=2\left ( \frac{K_{1}}{I_{1}}-\frac{K_{2}}{I_{2}}
  \right ) m_{\mathrm{s}},
\label{eq:alphaetc}  
\end{align}
where $\bar{\Sigma}_{\pi N}$ is related to the pion-nucleon
$\Sigma$ term: $\bar{\Sigma}_{\pi N} =(N_{c}-1)N_{c}^{-1}\Sigma_{\pi N}$.
As mentioned previously, the right hypercharge $\bar{Y}$ is fixed by
the number of valence quarks, i.e.,
$Y_R=(N_{c}-1)/3=2/3$. Diagonalizing the collective Hamiltonian, we
derive the collective wavefunctions of the singly heavy baryon 
\begin{align}
\psi_{B}^{(\mathcal{R})}(J',J'_{3};A) &= \sum_{m_{3}=\pm 1/2}
  C_{J_{Q}m_{3}JJ_{3}}^{J'J'_{3}}\sqrt{\mathrm{dim}(p,q)} 
(-1)^{-\frac{\bar{Y}}{2}+J_{3}}
  D_{(Y,T,T_{3}),(\bar{Y},J,-J_{3})}^{(\mathcal{R})*}(A) \chi_{m_3}, 
\label{eq:wf}
\end{align}
where $C_{J_{Q}m_{3}JJ_{3}}^{J'J'_{3}}$ denotes the Clebasch-Gordan
coefficient for the coupling between the collective light-quark
wavefunction and the heavy-quark spinor
$\chi_{m_3}$. $\mathrm{dim}(p,q)$ represents the dimension of the
$(p,q)$   
representation 
\begin{align}
&\mathrm{dim}(p,q) = (p+1)(q+1)\left(1+\frac{p+q}{2}\right).
\end{align}

In the presence of the flavor SU(3) symmetry breaking term
$H_{\mathrm{sb}}$, the collective wavefunctions of the baryon sextet            
should be mixed with those in higher representations. Thus, the
collective wavefunctions for the baryon antitriplet and sextet
are obtained respectively as 
\begin{align}
|B_{\bm{\overline{3}}_{0}}\rangle &= |\bm{\overline{3}}_{0},B\rangle 
  + p_{\overline{15}}^{B}|\bm{\overline{15}}_{0},B\rangle,
\label{eq:mixedWF1} \\
|B_{{\bm{6}}_{1}}\rangle &= |{\bm{6}}_{1},B\rangle 
  + q_{\overline{15}}^{B}|\bm{\overline{15}}_{1},B\rangle 
  + q_{\overline{24}}^{B}|\bm{\overline{24}}_{1},B\rangle
\label{eq:mixedWF2}
\end{align}
with the mixing coefficients
\begin{eqnarray}
p_{\overline{15}}^{B}
\;=\;
p_{\overline{15}} \left[\begin{array}{c}
-\sqrt{15}/10\\
-3\sqrt{5}/20
\end{array}\right], \quad
& 
q_{27}^{B}
\;=\; 
q_{\overline{15}}\left[\begin{array}{c}
\sqrt{5}/5\\
\sqrt{30}/20 \\
0
\end{array}\right], \quad
q_{\overline{24}}^{B}
\;=\;
q_{\overline{24}}\left[\begin{array}{c}
-\sqrt{10}/10\\
-\sqrt{15}/10 \\
-\sqrt{15}/10
\end{array}\right], 
\label{eq:pqmix1}
\end{eqnarray}
respectively, in the basis of 
$\left[\Lambda_{c}^{+},\;\Xi_{c} \right]$ for the baryon anti-triplet and
$\left[\Sigma_{c}(\Sigma_{c}^{*}),\;\Xi_{c}^{\prime}(\Xi_{c}^{*})
,\;\Omega_{c}^{0}(\Omega_{c}^{*0})\right]$ for the baryon sextet. 
The parameters $p_{\overline{15}}$, $q_{\overline{15}}$ and
$q_{\overline{24}}$ are given in terms of the inertia parameters
$\alpha$ and $\gamma$
\begin{align}
p_{\overline{15}} =\frac{3}{4\sqrt{3}} \alpha I_{2}, \quad 
q_{\overline{15}} = -\frac{1}{\sqrt{2}} \left( \alpha +
  \frac{2}{3}\gamma \right)I_{2},\quad  
q_{\overline{24}} = \frac{4}{5\sqrt{10}} \left( \alpha -
  \frac{1}{3}\gamma \right)I_{2}. 
\label{eq:pqmix2}
\end{align}

It is straightforward to compute the transition matrix elements of the
collective states, which will be expressed by the SU(3) Clebsch-Gordan
coefficients. So, we arrive at the final expressions for the
axial-vector transition form factors of the singly heavy baryon with
spin 1/2 and 3/2 respectively 
\begin{align}
G_{A,\, B \rightarrow B'}^{(\chi)}(Q^{2}) =& \sqrt{2}
  \left[ \frac{ \langle D^{(8)}_{a3} \rangle}{3} 
  \{\mathcal{A}^{B \rightarrow B'}_{0}(Q^{2}) 
  -\mathcal{A}^{B \rightarrow B'}_{2}(Q^{2})\}
  -\frac{i \langle D^{(8)}_{a3} \rangle}{6I_{1}} 
  \{\mathcal{D}^{B \rightarrow B'}_{0}(Q^{2})
  -\mathcal{D}^{B \rightarrow B'}_{2}(Q^{2})\} \right. \cr
&+\frac{1}{3\sqrt{3} I_{1}} \left[ \langle D^{(8)}_{a8} \hat{J}_{3} 
  \rangle +\frac{2m_{\mathrm{s}}}{\sqrt{3}} K_{1} \langle D^{(8)}_{83}
  D^{(8)}_{a8} \rangle \right] 
  \{\mathcal{B}^{B \rightarrow B'}_{0}(Q^{2})
  -\mathcal{B}^{B \rightarrow B'}_{2}(Q^{2})\} \cr 
& +\frac{d_{pq3}}{3 I_{2}} \left[ \langle D^{(8)}_{ap} \hat{J}_{q} 
  \rangle +\frac{2m_{\mathrm{s}}}{\sqrt{3}} K_{2} \langle D^{(8)}_{ap} 
  D^{(8)}_{8q} \rangle \right]
  \{\mathcal{C}^{B \rightarrow B'}_{0}(Q^{2}) 
  -\mathcal{C}^{B \rightarrow B'}_{2}(Q^{2})\} \cr 
& +\frac{2 m_{\mathrm{s}}}{9} ( \langle D^{(8)}_{a3} \rangle - \langle 
  D^{(8)}_{88}D^{(8)}_{a3} \rangle) \{\mathcal{H}^{B\rightarrow B'}_{0}(Q^{2})
  -\mathcal{H}^{B \rightarrow B'}_{2}(Q^{2})\} \cr 
& -\frac{2 m_{\mathrm{s}}}{9} 
  \langle D^{(8)}_{83} D^{(8)}_{a8} \rangle 
  \{\mathcal{I}^{B \rightarrow B'}_{0}(Q^{2})
  -\mathcal{I}^{B \rightarrow B'}_{2}(Q^{2})\} \cr 
& \left. -\frac{2 m_{\mathrm{s}}}{3\sqrt{3}} d_{pq3} \langle D^{(8)}_{ap} 
  D^{(8)}_{8q} \rangle \{\mathcal{J}^{B \rightarrow B'}_{0}(Q^{2})
  -\mathcal{J}^{B \rightarrow B'}_{2}(Q^{2})\} \right],
\label{eq:GAform} \\
C_{5,\, B \rightarrow B'}^{A(\chi)}(Q^{2}) =& \sqrt{3}
  \left[ \frac{ \langle D^{(8)}_{a3} \rangle}{3} 
  \{\mathcal{A}^{B \rightarrow B'}_{0}(Q^{2}) 
  -\mathcal{A}^{B \rightarrow B'}_{2}(Q^{2})\}
  -\frac{i \langle D^{(8)}_{a3} \rangle}{6I_{1}} 
  \{\mathcal{D}^{B \rightarrow B'}_{0}(Q^{2})
  -\mathcal{D}^{B \rightarrow B'}_{2}(Q^{2})\} \right. \cr
&+\frac{1}{3\sqrt{3} I_{1}} \left[ \langle D^{(8)}_{a8} \hat{J}_{3} 
  \rangle +\frac{2m_{\mathrm{s}}}{\sqrt{3}} K_{1} \langle D^{(8)}_{83}
  D^{(8)}_{a8} \rangle \right] 
  \{\mathcal{B}^{B \rightarrow B'}_{0}(Q^{2})
  -\mathcal{B}^{B \rightarrow B'}_{2}(Q^{2})\} \cr 
& +\frac{d_{pq3}}{3 I_{2}} \left[ \langle D^{(8)}_{ap} \hat{J}_{q} 
  \rangle +\frac{2m_{\mathrm{s}}}{\sqrt{3}} K_{2} \langle D^{(8)}_{ap} 
  D^{(8)}_{8q} \rangle \right]
  \{\mathcal{C}^{B \rightarrow B'}_{0}(Q^{2}) 
  -\mathcal{C}^{B \rightarrow B'}_{2}(Q^{2})\} \cr 
& +\frac{2 m_{\mathrm{s}}}{9} ( \langle D^{(8)}_{a3} \rangle - \langle 
  D^{(8)}_{88}D^{(8)}_{a3} \rangle) \{\mathcal{H}^{B\rightarrow B'}_{0}(Q^{2})
  -\mathcal{H}^{B \rightarrow B'}_{2}(Q^{2})\} \cr 
& -\frac{2 m_{\mathrm{s}}}{9} 
  \langle D^{(8)}_{83} D^{(8)}_{a8} \rangle 
  \{\mathcal{I}^{B \rightarrow B'}_{0}(Q^{2})
  -\mathcal{I}^{B \rightarrow B'}_{2}(Q^{2})\} \cr 
& \left. -\frac{2 m_{\mathrm{s}}}{3\sqrt{3}} d_{pq3} \langle D^{(8)}_{ap} 
  D^{(8)}_{8q} \rangle \{\mathcal{J}^{B \rightarrow B'}_{0}(Q^{2})
  -\mathcal{J}^{B \rightarrow B'}_{2}(Q^{2})\} \right],
\label{eq:C5form} 
\end{align}
where $\langle \cdots \rangle$ designate the transition baryonic
matrix elements of given collective operators. The explicit
expressions for the quark densities 
$\mathcal{A}_{0,2}$, $\mathcal{B}_{0,2}$, $\mathcal{C}_{0,2}$,
$\mathcal{D}_{0,2}$, $\mathcal{H}_{0,2}$, $\mathcal{I}_{0,2}$, and
$\mathcal{J}_{0,2}$ can be found in Appendix~\ref{app:a}. 
Note that the corrections from flavor SU(3) symmetry breaking are
originated from two different sources: that from the effective chiral
action and that from the collective wavefunctions, which we denote
them respectively as $(G_{A,\, B \rightarrow B'}^{(\chi)}
)^{(\mathrm{op})}$ and $(G_{A,\, B \rightarrow B'}^{(\chi)})^{(\mathrm{wf})}$ 
\begin{align}
G_{A,\, B \rightarrow B'}^{(\chi)}(Q^{2}) &= (G_{A,\, B \rightarrow
                                         B'}^{(\chi)}
                                         (Q^{2}))^{(\mathrm{sym})}  
  +(G_{A,\, B \rightarrow B'}^{(\chi)} (Q^{2}))^{(\mathrm{op})} 
  +(G_{A,\, B \rightarrow B'}^{(\chi)} (Q^{2}))^{(\mathrm{wf})}. 
\label{eq:GA_decompose}
\end{align}
We decompose $C_{5,\, B \rightarrow B'}^{A(\chi)}$ in the same manner 
\begin{align}
C_{5,\, B \rightarrow B'}^{A(\chi)}(Q^{2}) &= (C_{5,\, B \rightarrow
                                          B'}^{A(\chi)}
                                          (Q^{2}))^{(\mathrm{sym})}  
  +(C_{5,\, B \rightarrow B'}^{A(\chi)} (Q^{2}))^{(\mathrm{op})} 
  +(C_{5,\, B \rightarrow B'}^{A(\chi)} (Q^{2}))^{(\mathrm{wf})}. 
\label{eq:C5_decompose}
\end{align}
Having computed the transition matrix elements of the $D$ functions,
we find the following results respectively for $G_{A,\, B \rightarrow
  B'}^{(\chi)}(Q^2)^{(\mathrm{sym})}$, 
and $G_{A,\, B \rightarrow B'}^{(\chi)}(Q^2)^{(\mathrm{op})}$, and
$G_{A,\, B \rightarrow B'}^{(\chi)}(Q^2)^{(\mathrm{wf})}$
\begin{align}
G_{A,\, B \rightarrow B'}^{(\chi)}(Q^{2})^{(\mathrm{sym})} &= -\frac{\sqrt{3}}{36} 
\left( \begin{array}{c} \sqrt{2} \\ -\sqrt{2}T_{3} \\ -\sqrt{2} \\
           1 \\ -\sqrt{2} \end{array} \right)
  \left[  \left\{ 2\{\mathcal{A}^{B \rightarrow B'}_{0}(Q^{2})
  -\mathcal{A}^{B \rightarrow B'}_{2}(Q^{2})\} 
  -\frac{i\{\mathcal{D}^{B \rightarrow B'}_{0}(Q^{2})
 -\mathcal{D}^{B \rightarrow B'}_{2}(Q^{2})\}}{I_{1}}\right\} \right.\cr
&\left. \hspace{3.5cm} - \frac{\{\mathcal{C}^{B \rightarrow B'}_{0}(Q^{2})
  -\mathcal{C}^{B \rightarrow B'}_{2}(Q^{2})\}}{I_{2}} \right] ,  
\label{eq:GA3sym} \\
G_{A,\, B \rightarrow B'}^{(\chi)}(Q^{2})^{(\mathrm{op})} =& 
-\frac{\sqrt{3}m_{\mathrm{s}}}{540} 
  \Bigg[ \sqrt{2}\left( \begin{array}{c} \sqrt{2} 
\\ -2\sqrt{2}T_{3} \\ \sqrt{2} \\
   -3 \\ 0 \end{array} \right) 
  \left\{ \frac{K_{1}}{I_{1}} \{\mathcal{B}^{B \rightarrow B'}_{0}(Q^{2}) 
   -\mathcal{B}^{B \rightarrow B'}_{2}(Q^{2})\} 
-\{\mathcal{I}^{B \rightarrow B'}_{0}
  (Q^{2}) -\mathcal{I}^{B \rightarrow B'}_{2}(Q^{2})\} \right\} \cr 
& \hspace{1.0cm} +3\sqrt{2}\left( \begin{array}{c} 2\sqrt{2} 
\\ -2\sqrt{2}T_{3} \\ -\sqrt{2} \\ 1 \\ -\sqrt{2} \end{array} 
  \right) \left\{ \frac{K_{2}}{I_{2}} 
  \{\mathcal{C}^{B \rightarrow B'}_{0}(Q^{2})-\mathcal{C}^{B
  \rightarrow B'}_{2}(Q^{2})\} 
-\{\mathcal{J}^{B \rightarrow B'}_{0}(Q^{2}) 
  -\mathcal{J}^{B \rightarrow B'}_{2}(Q^{2})\} \right\} \cr 
& \hspace{1.0cm} + \left( \begin{array}{c} 7\sqrt{2} 
\\ -10\sqrt{2}T_{3} \\ -11\sqrt{2} \\ 9 \\ -12\sqrt{2} 
  \end{array} \right) \{\mathcal{H}^{B \rightarrow B'}_{0}(Q^{2}) 
  -\mathcal{H}^{B \rightarrow B'}_{2}(Q^{2})\} \Bigg], 
\label{eq:GA3opcorr} 
\end{align}
\begin{align}
G_{A,\, B \rightarrow B'}^{(\chi)}(Q^{2})^{(\mathrm{wf})} =& 
   -\frac{\sqrt{2}}{1440}p_{\overline{15}}\left( \begin{array}{c}
 2 \\ -5T_{3} \\ -1 \\ -3\sqrt{2} \\ 3  
  \end{array} \right)\Bigg[\sqrt{2}  
\left\{ 2\{\mathcal{A}^{B \rightarrow B'}_{0}(Q^{2})
  -\mathcal{A}^{B \rightarrow B'}_{2}(Q^{2})\} 
  \right. \cr&\left.\hspace{4.0cm}
  -\frac{i\{\mathcal{D}^{B \rightarrow B'}_{0}(Q^{2})-\mathcal{D}^{B
  \rightarrow B'}_{2}(Q^{2})\}}{I_{1}} \right\} \cr 
& \hspace{3.5cm}  + 3
  \frac{\{\mathcal{C}^{B \rightarrow B'}_{0}(Q^{2})-\mathcal{C}^{B
  \rightarrow B'}_{2}(Q^{2})\}}{I_{2}} \Bigg]\cr 
&  -\frac{\sqrt{30}}{7200} q_{\overline{15}}\Bigg[ 
  \left( \begin{array}{c} 4 \\ -T_{3} \\  8 \\ -3\sqrt{2} \\ 0 
  \end{array} \right)
   \left\{ 2\{\mathcal{A}^{B \rightarrow B'}_{0}(Q^{2})
  -\mathcal{A}^{B \rightarrow B'}_{2}(Q^{2})\} 
  \right.\cr &\left.\hspace{4.0cm}
  -\frac{i\{\mathcal{D}^{B \rightarrow B'}_{0}(Q^{2})
  -\mathcal{D}^{B \rightarrow B'}_{2}(Q^{2})\}}{I_{1}}\right\} \cr
& \hspace{1.5cm}  -2\sqrt{5}\left( \begin{array}{c} 4\sqrt{2} 
\\ -\sqrt{2}T_{3} \\ 4\sqrt{2} \\ -3 \\ 0 
  \end{array} \right)\frac{\{\mathcal{C}^{B \rightarrow B'}_{0}(Q^{2})
  -\mathcal{C}^{B \rightarrow B'}_{2}(Q^{2})\}}{I_{2}} \Bigg],
\label{eq:GA3wfcorr}
\end{align}
in the basis of $\left[\Sigma_{c}^{+} \rightarrow \Lambda_{c}^{+},\;
  \Xi_{c}^{\prime }\rightarrow \Xi_{c},\; 
\Sigma_{c}^{++}\rightarrow \Xi_{c}^{+}, \; \Xi_{c}^{\prime 0}
\rightarrow \Lambda_{c}^{+},\; \Omega_{c}^{0} \rightarrow \Xi_{c}^{0} 
 \right]$, and respectively for
$C_{5,\, B \rightarrow B'}^{A(\chi)}(Q^2)^{(\mathrm{sym})}$, 
$C_{5,\, B \rightarrow B'}^{A(\chi)}(Q^2)^{(\mathrm{op})}$, and 
$C_{5,\, B \rightarrow B'}^{A(\chi)}(Q^2)^{(\mathrm{wf})}$  
\begin{align}
C_{5,\, B \rightarrow B'}^{A(\chi)}(Q^{2})^{(\mathrm{sym})} &= -\frac{1}{90} 
  \left( \begin{array}{c} T_{3} \\ T_{3} \\ 1 \\ 1 \end{array} \right)
  \left[3\left\{ 2\{\mathcal{A}^{B \rightarrow B'}_{0}(Q^{2})
  -\mathcal{A}^{B \rightarrow B'}_{2}(Q^{2})\} 
  -\frac{i\{\mathcal{D}^{B \rightarrow B'}_{0}(Q^{2})
 -\mathcal{D}^{B \rightarrow B'}_{2}(Q^{2})\}}{I_{1}}\right\} \right.\cr
& \left.\hspace{2.5cm} - 3\frac{\{\mathcal{C}^{B \rightarrow B'}_{0}(Q^{2})
  -\mathcal{C}^{B \rightarrow B'}_{2}(Q^{2})\}}{I_{2}} 
  - 2\frac{\{\mathcal{B}^{B \rightarrow B'}_{0}(Q^{2})
  -\mathcal{B}^{B \rightarrow B'}_{2}(Q^{2})\}}{I_{1}}\right]  ,  
\label{eq:C53sym} 
\end{align}
\begin{align}
C_{5,\, B \rightarrow B'}^{A(\chi)}(Q^{2})^{(\mathrm{op})} =&
 -\frac{2\sqrt{2}m_{\mathrm{s}}}{810}  
  \Bigg[ \sqrt{2}\left( \begin{array}{c} 4T_{3} \\ T_{3} \\ -1 \\ -3 \end{array} \right) 
  \left\{ \frac{K_{1}}{I_{1}} \{\mathcal{B}^{B \rightarrow B'}_{0}(Q^{2}) 
   -\mathcal{B}^{B \rightarrow B'}_{2}(Q^{2})\} 
-\{\mathcal{I}^{B \rightarrow B'}_{0}
  (Q^{2}) -\mathcal{I}^{B \rightarrow B'}_{2}(Q^{2})\} \right\} \cr 
& \hspace{1.0cm} +\sqrt{2}\left( \begin{array}{c} 10T_{3} \\ 14T_{3} \\ -7 \\ -3 \end{array} 
  \right) \left\{ \frac{K_{2}}{I_{2}} 
  \{\mathcal{C}^{B \rightarrow B'}_{0}(Q^{2})-\mathcal{C}^{B
  \rightarrow B'}_{2}(Q^{2})\} 
-\{\mathcal{J}^{B \rightarrow B'}_{0}(Q^{2}) 
  -\mathcal{J}^{B \rightarrow B'}_{2}(Q^{2})\} \right\} \cr 
& \hspace{1.0cm} + \left( \begin{array}{c} 14\sqrt{2}T_{3} \\ 16\sqrt{2}T_{3} \\ 19\sqrt{2} \\ 21 
  \end{array} \right) \{\mathcal{H}^{B \rightarrow B'}_{0}(Q^{2}) 
  -\mathcal{H}^{B \rightarrow B'}_{2}(Q^{2})\} \Bigg], 
\label{eq:C53opcorr} 
\end{align}
\begin{align}
C_{5,\, B \rightarrow B'}^{A(\chi)}(Q^{2})^{(\mathrm{wf})} =& 
   -\frac{1}{90}q_{\overline{15}}\Bigg[ 2\sqrt{5}\left( \begin{array}{c}
  8T_{3} \\  4T_{3} \\  -5 \\ -3  
  \end{array} \right) \left\{ 2\{\mathcal{A}^{B \rightarrow B'}_{0}(Q^{2})
  -\mathcal{A}^{B \rightarrow B'}_{2}(Q^{2})\} 
  -\frac{i\{\mathcal{D}^{B \rightarrow B'}_{0}(Q^{2})-\mathcal{D}^{B
  \rightarrow B'}_{2}(Q^{2})\}}{I_{1}} \right\} \cr 
& \hspace{1.2cm}  + \sqrt{5}\left( \begin{array}{c} 8T_{3} \\ 5T_{3} \\
                              -5 \\ 3
  \end{array} \right) 
  \frac{\{\mathcal{C}^{B \rightarrow B'}_{0}(Q^{2})-\mathcal{C}^{B
  \rightarrow B'}_{2}(Q^{2})\}}{I_{2}} \cr
  &\hspace{1.2cm}-2 \left( \begin{array}{c} 6\sqrt{6}T_{3} \\ 15T_{3} \\
                              5\sqrt{15} \\ 3\sqrt{15}
  \end{array} \right)
  \frac{\{\mathcal{B}^{B \rightarrow B'}_{0}(Q^{2})-\mathcal{B}^{B
  \rightarrow B'}_{2}(Q^{2})\}}{I_{1}} \Bigg] \cr
  & +\frac{\sqrt{5}}{5400}q_{\overline{24}}\Bigg[ \sqrt{2}\left( \begin{array}{c}
  0 \\  0 \\  2 \\ -1  
  \end{array} \right) \left\{ 2\{\mathcal{A}^{B \rightarrow B'}_{0}(Q^{2})
  -\mathcal{A}^{B \rightarrow B'}_{2}(Q^{2})\} 
  -\frac{i\{\mathcal{D}^{B \rightarrow B'}_{0}(Q^{2})-\mathcal{D}^{B
  \rightarrow B'}_{2}(Q^{2})\}}{I_{1}} \right\} \cr 
& \hspace{1.2cm}  + 8\sqrt{2}\left( \begin{array}{c} 0 \\ 0 \\
                              1 \\ 0
  \end{array} \right) 
  \frac{\{\mathcal{C}^{B \rightarrow B'}_{0}(Q^{2})-\mathcal{C}^{B
  \rightarrow B'}_{2}(Q^{2})\}}{I_{2}} \cr
  &\hspace{1.2cm}+40 \left( \begin{array}{c} 0 \\ 0 \\
                              1 \\ 0
  \end{array} \right)
  \frac{\{\mathcal{B}^{B \rightarrow B'}_{0}(Q^{2})-\mathcal{B}^{B
  \rightarrow B'}_{2}(Q^{2})\}}{I_{1}} \Bigg] 
\label{eq:C53wfcorr}
\end{align}
in the basis of $\left[\Sigma_{c}^{*} \rightarrow \Sigma_{c},\;
  \Xi_{c}^{* } \rightarrow \Xi_{c}^{\prime },\;
  \Xi_{c}^{*+}\rightarrow \Sigma_{c}^{++},\; \Omega_{c}^{*
    0}\rightarrow \Xi_{c}^{\prime +}\right]$. 
The explicit expressions for $\mathcal{A}_i^{B\to B'}$,
$\mathcal{B}_i^{B\to B'}$, $\mathcal{C}_i^{B\to B'}$,
$\mathcal{D}_i^{B\to B'}$, $\mathcal{I}_i^{B\to B'}$,
$\mathcal{J}_i^{B\to B'}$, and $\mathcal{H}_i^{B\to B'}$ can be found
in Appendix~\ref{app:a}. The matrix elements of the collective
operators are given in Appendix~\ref{app:b} in detail.

Scrutinizing Eqs.~\eqref{eq:GA3sym}-\eqref{eq:C53wfcorr}, we find
isospin relations between different axial-vector transition form
factors,
\begin{align}
(\Xi_{c}^{\prime +} \rightarrow \Xi_{c}^{+}) &= -(\Xi_{c}^{\prime 0}
  \rightarrow  \Xi_{c}^{0})  
= \frac{1}{\sqrt{2}}(\Xi_{c}^{\prime +} \rightarrow \Xi_{c}^{0}) 
= -\frac{1}{\sqrt{2}}(\Xi_{c}^{\prime 0} \rightarrow \Xi_{c}^{+})\cr
(\Sigma_{c}^{+} \rightarrow \Lambda_{c}^{+})& = (\Sigma_{c}^{++}
   \rightarrow   \Lambda_{c}^{+}) 
= (\Sigma_{c}^{0} \rightarrow \Lambda_{c}^{+}) \cr
(\Xi_{c}^{\prime +} \rightarrow \Lambda_{c}^{+}) &= (\Xi_{c}^{\prime
   0} \rightarrow   \Lambda_{c}^{+})   \cr 
(\Sigma_{c}^{++} \rightarrow \Xi_{c}^{+}) &= (\Sigma_{c}^{0}
   \rightarrow \Xi_{c}^{0}) 
= \sqrt{2}(\Sigma_{c}^{+} \rightarrow \Xi_{c}^{0}) =
  \sqrt{2}(\Sigma_{c}^{+}  \rightarrow \Xi_{c}^{+})  \cr 
(\Omega_{c}^{0} \rightarrow \Xi_{c}^{+}) &= -(\Omega_{c}^{0}
  \rightarrow \Xi_{c}^{0})  \cr 
(\Sigma_{c}^{* ++} \rightarrow \Sigma_{c}^{++}) &= -(\Sigma_{c}^{* 0}
  \rightarrow  \Sigma_{c}^{0}) 
= -(\Sigma^{* ++} \rightarrow \Sigma_{c}^{+}) = 
(\Sigma_{c}^{* +} \rightarrow \Sigma_{c}^{++}) \cr
&= -(\Sigma^{* +} \rightarrow \Sigma_{c}^{0})= 
(\Sigma_{c}^{* 0} \rightarrow \Sigma_{c}^{+}) \cr 
(\Xi_{c}^{* +} \rightarrow \Xi_{c}^{\prime +})& = 
-(\Xi_{c}^{* 0} \rightarrow \Xi_{c}^{\prime 0}) 
= \frac{1}{\sqrt{2}}(\Xi_{c}^{* 0} \rightarrow \Xi_{c}^{\prime +})
= -\frac{1}{\sqrt{2}}(\Xi_{c}^{* +} \rightarrow \Xi_{c}^{\prime 0})\cr 
(\Xi_{c}^{* 0} \rightarrow \Sigma_{c}^{+}) &=
-(\Sigma^{* +} \rightarrow \Xi_{c}^{\prime 0}) 
= \frac{1}{\sqrt{2}}(\Xi_{c}^{* +} \rightarrow \Sigma_{c}^{++}) 
= -\frac{1}{\sqrt{2}}(\Sigma_{c}^{* ++} \rightarrow \Xi_{c}^{\prime +}) 
= -(\Xi_{c}^{* +} \rightarrow \Sigma_{c}^{+}) \cr
& = -(\Sigma^{* +} \rightarrow \Xi_{c}^{\prime +})
= -\frac{1}{\sqrt{2}}(\Sigma^{* 0} \rightarrow \Xi_{c}^{\prime +})
= \frac{1}{\sqrt{2}}(\Xi_{c}^{* 0} \rightarrow \Sigma_{c}^{0}) \cr
(\Omega_{c}^{* 0} \rightarrow \Xi_{c}^{\prime +}) &= -(\Xi_{c}^{* +}
  \rightarrow  \Omega_{c}^{0})  
= (\Omega_{c}^{* 0} \rightarrow \Xi_{c}^{\prime 0}) = -(\Xi_{c}^{* 0+}
  \rightarrow  \Omega_{c}^{0}),
  \label{eq:relations1}
  \end{align}
 the SU(3) symmetric relations
  \begin{align}
(\Sigma_{c}^{+} \rightarrow \Lambda_{c}^{+})&
 =-2(\Xi_{c}^{\prime +} \rightarrow \Xi_{c}^{+}) 
= -(\Sigma_{c}^{++} \rightarrow \Xi_{c}^{+}) 
=\sqrt{2}(\Xi_{c}^{\prime 0} \rightarrow \Lambda_{c}^{+})
= -(\Omega_{c}^{0} \rightarrow \Xi_{c}^{0}) \cr
(\Sigma_{c}^{* ++} \rightarrow \Sigma_{c}^{++}) 
&=2(\Xi_{c}^{* +} \rightarrow \Xi_{c}^{\prime +})
= (\Xi_{c}^{* +} \rightarrow \Sigma_{c}^{++})
=(\Omega_{c}^{* 0} \rightarrow \Xi_{c}^{\prime +}),
\label{eq:relations2}
  \end{align}
and the various sum rules as follows
  \begin{align}
(\Sigma^{* +} \rightarrow \Sigma_{c}^{++}) &= 
(\Xi_{c}^{* +} \rightarrow \Omega_{c}^{0})
+(\Xi_{c}^{* +} \rightarrow \Sigma_{c}^{++})
-\sqrt{2}(\Xi_{c}^{* +} \rightarrow \Xi_{c}^{\prime 0}) \cr
(\Sigma^{* +} \rightarrow \Xi_{c}^{0})&= -\frac{1}{2}
(\Xi_{c}^{\prime 0} \rightarrow \Lambda_{c}^{+})
+\frac{1}{\sqrt{2}}(\Xi_{c}^{* +} \rightarrow \Xi_{c}^{+})
+\frac{\sqrt{3}}{2\sqrt{2}}(\Xi_{c}^{\prime +} \rightarrow \Omega_{c}^{0}) \cr
&+\frac{1}{4\sqrt{6}}(\Xi_{c}^{* +} \rightarrow \Sigma_{c}^{++})
-\frac{5}{4\sqrt{3}}(\Xi_{c}^{* +} \rightarrow \Xi_{c}^{\prime 0}) \cr 
(\Sigma^{* +} \rightarrow \Lambda_{c}^{+})&= \frac{3}{2\sqrt{2}}
(\Xi_{c}^{\prime 0} \rightarrow \Lambda_{c}^{+})
-\frac{1}{2}(\Xi_{c}^{* +} \rightarrow \Xi_{c}^{+})
+\frac{3\sqrt{3}}{4}(\Xi_{c}^{\prime +} \rightarrow \Omega_{c}^{0}) \cr
&+\frac{\sqrt{3}}{8}(\Xi_{c}^{* +} \rightarrow \Sigma_{c}^{++})
-\frac{5\sqrt{3}}{4\sqrt{2}}(\Xi_{c}^{* +} \rightarrow \Xi_{c}^{\prime 0}) \cr
(\Omega_{c}^{* 0} \rightarrow \Xi_{c}^{+}) &= 
-\frac{1}{2\sqrt{2}}(\Xi_{c}^{\prime 0} \rightarrow \Lambda_{c}^{+})
+\frac{3}{2}(\Xi_{c}^{* +} \rightarrow \Xi_{c}^{+})
+\frac{3\sqrt{3}}{4}(\Xi_{c}^{\prime +} \rightarrow \Omega_{c}^{0}) \cr
&+\frac{\sqrt{3}}{8}(\Xi_{c}^{* +} \rightarrow \Sigma_{c}^{++})
-\frac{5\sqrt{3}}{4\sqrt{2}}(\Xi_{c}^{* +} \rightarrow \Xi_{c}^{\prime 0}). 
\label{eq:relations}
\end{align}
These relations indicate that not all form factors are independent. 
As we will show soon, we will only have two different form factors,
from which all other form factors can be easily obtained when flavor
SU(3) symmetry is imposed. 

While there are no experimental information on the axial-vector
transition form factors of the singly heavy baryons, their strong decay
widths are experimentally known. In particular, the Belle
Collaboration has recently reported those for the $\Sigma_c^+\to
\Lambda_c^+ + \pi^0$ and $\Sigma_c^{*+}\to \Lambda_c^+ + \pi^0$
decays~\cite{Belle:2021qip}. Thus, we will also compute all possible
strong decay widths for the singly heavy baryons, using the following
formulae 
\begin{align}
&\Gamma_{B_{1/2} \rightarrow B_{1/2}m_{\pi}} = 
\frac{1}{8\pi }\frac{|\bm{q}|^{3}}{f_{\pi}^{2}}
\frac{M_{f}}{M_{i}}(G_{A,\, B \rightarrow B' }(0))^{2},\cr
&\Gamma_{B_{3/2} \rightarrow B_{1/2}m_{\pi}} = 
\frac{1}{12\pi }\frac{|\bm{q}|^{3}}{f_{\pi}^{2}}
\frac{M_{f}}{M_{i}}(C_{5,\, B \rightarrow B'}^{A }(0))^{2}, 
\label{eq:decay_w}
\end{align}
where the pion momentum $|\bm{q}|$ is given by
\begin{align}
|\bm{q}| =
  \frac{1}{2M_{i}}\sqrt{\left(M_{i}^{2}-(M_{f}^{2}+m_{8})^{2}
\right)\left(M_{i}^{2}-(M_{f}^{2}-m_{8})^{2}\right)}.
\end{align}
$M_{i}$, $M_{f}$ are the initial and final baryon masses,
respectively. $m_{\pi}$ represents the mass of the pion and $f_{\pi}$
stands for the pion decay constant. The mass ratio $M_f/M_i$ in
Eq.~\eqref{eq:decay_w} arises from the recoil
effect~\cite{Cheng:2006dk, Cheng:2015naa}. Since the velocities of
$B_i$ and $B_f$ are the same to order $\mathcal{O}(1/m_Q)$ in
effective heavy quark theory, the recoil effects can be given by the
mass ratio.  

\section{Results and discussion}
\label{sec:5}
We now show the results for the axial-vector transition form factors
of the singly heavy baryons and discuss them. We first explain how the
model parameters are fixed. In the $\chi$QSM, four different
parameters need to be determined: the dynamical quark mass $M$, the
cutoff mass $\Lambda$ in the regularization functions, the strange
current quark mass $m_{\mathrm{s}}$, and the average of the up and
down current quarks $\overline{m}$, as mentioned in Section
III. $\overline{m}$ is determined by reproducing the physical value of
the pion mass, $m_\pi=140$ MeV. The strange current quark mass is
usually fixed by the kaon mass, $m_K = 495$ MeV. We use
$m_{\mathrm{s}}=180$ MeV, which reproduces approximately the kaon mass
and describes the mass spectra of the baryon octet and
decuplet~\cite{Blotz:1992pw, Christov:1995vm} very well. The 
cutoff mass $\Lambda$ is fixed by the pion decay constant
$f_\pi=93$ MeV. The dynamical quark mass $M$ is regarded as a free
parameter in the $\chi$QSM. Nevertheless, we use $M=420$ MeV because
one can produce various experimental data such as the radius of the
proton~\cite{Kim:1995mr}, the magnetic dipole
moments~\cite{Kim:1997ip}, and semileptonic decays of
hyperons~\cite{Kim:1997ts, Ledwig:2008ku}. Note that the values of all
the parameters are the same as in the previous
works~\cite{Kim:2018xlc, Kim:2018nqf, Kim:2020uqo, Suh:2022guw}.  

Since we use the $1/N_c$ and $1/m_Q$ expansion as a guiding principle,  
we have to maintain consistency in dealing with its expansion within
the theoretical framework. So, we can ignore the mass difference
in the momentum transfer. It indicates that momentum transfer in
Eq.~\eqref{eq:Momenta} can be approximated to be $\bm{q}^2\approx
-Q^2$. The expressions for $G_A^{(\chi)}$ and $C_5^{A(\chi)}$ contain
the masses of the singly heavy baryons. Keeping the $1/N_c$ expansion
in mind, we approximate $M_i$ and $M_f$ by
$M_{\mathrm{cl}}+m_{\mathrm{c}}$. The baryon masses in the $\chi$QSM
also include the rotational $1/N_c$ and $m_{\mathrm{s}}$
corrections. If we turn off all the corrections, the singly heavy
baryon mass becomes the classical $N_c-1$ soliton mass
$M_{\mathrm{cl}}$ plus the charm quark mass. To be theoretically more
consistent, hence, we will take $M_{\mathrm{cl}}+m_{\mathrm{c}}$
instead of a antitriplet and sextet baryon masses. In effect, the
numerical results are improved by considering
$M_{\mathrm{cl}}+m_{\mathrm{c}}$ in place of $M_{\bar{3}}$ and $M_{6}$
by around 10~\%. Similar approximations were performed in the case of
the baryon octet and decuplet~\cite{Meissner:1986js, Ledwig:2008es,
  Suh:2022guw}. 

\begin{figure}[htp]
\centering
  \includegraphics[scale=0.54]{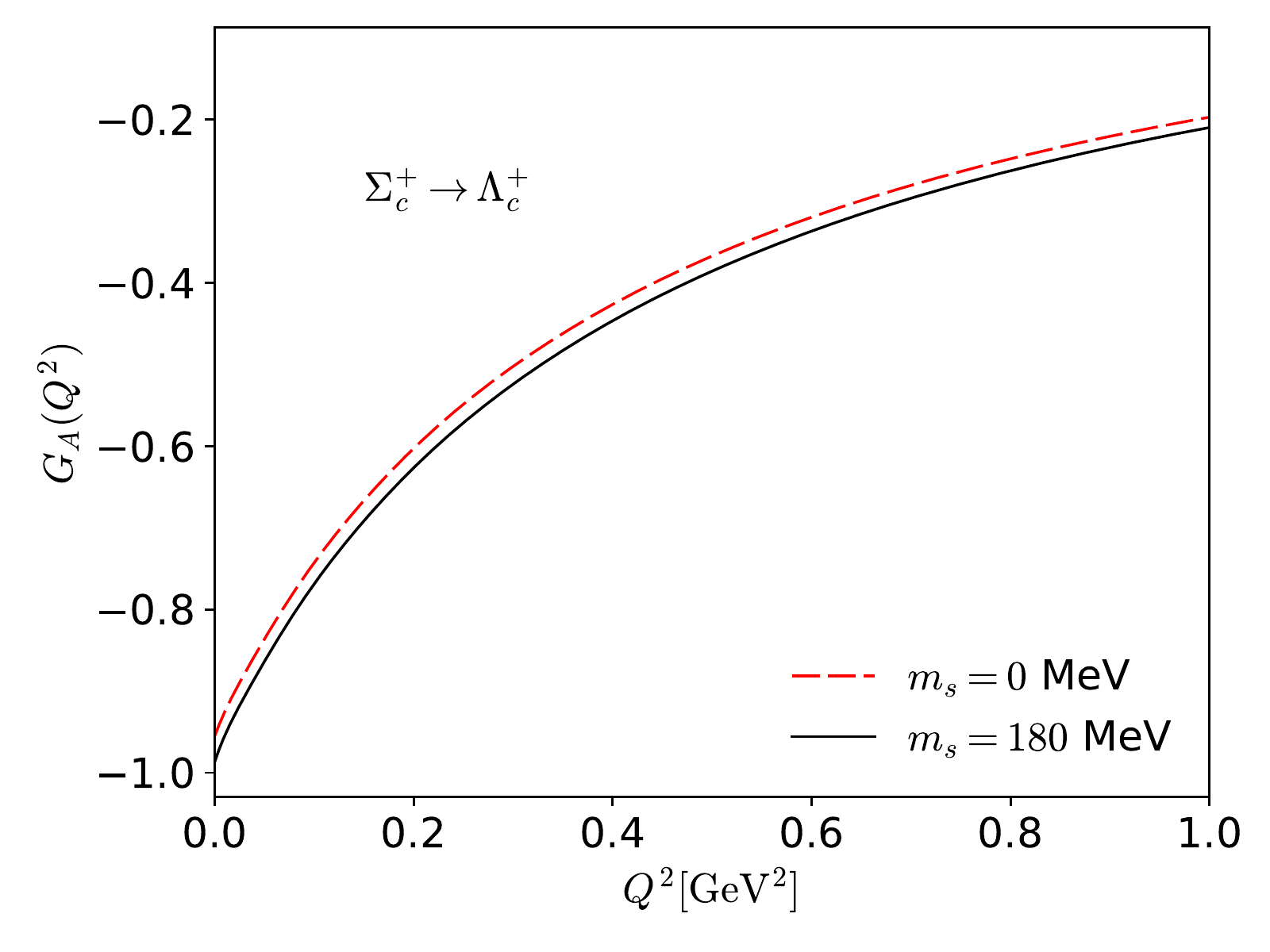}
  \includegraphics[scale=0.54]{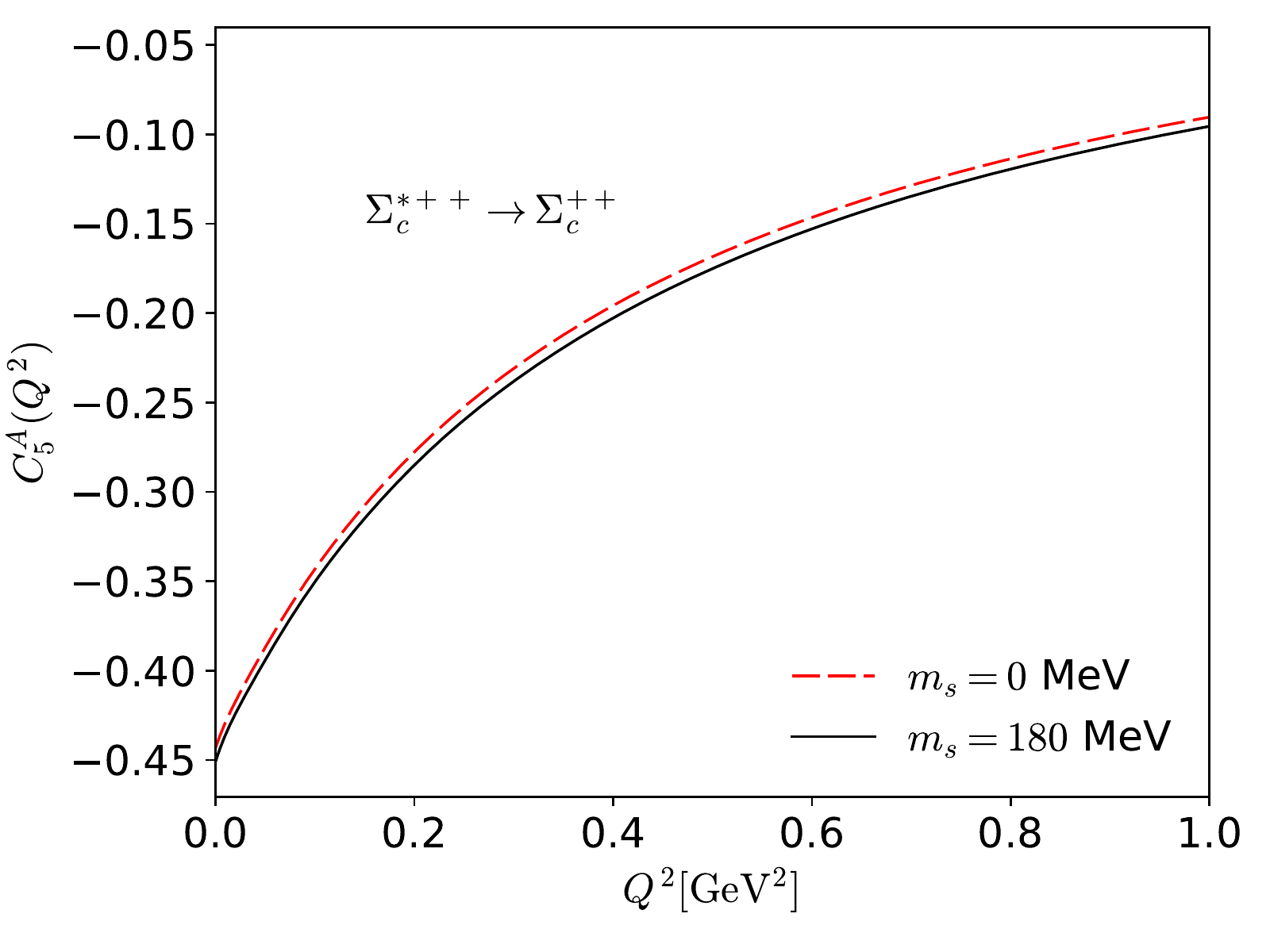}
\caption{The axial-vector transition form factor $G_{A,\, \Sigma_c^+
    \rightarrow \Lambda_c^+}^{(3)}(Q^{2})$ ($C_{5, \; \Sigma_c^{*++}
    \rightarrow \Sigma_c^{++}}^{A(3)}(Q^{2})$) for 
the transitions from the baryon sextet with spin-$1/2$ ($3/2$) to the
baryon anti-triplet.  
In the left panel, the form factors for the $\Sigma_c^+\to \Lambda_c^+$
transition are drawn, whereas in the right panel, those for the
$\Sigma_c^{*++} \to \Sigma_c^{++}$ are depicted. The solid and dashed
curves represent the total results with the effects of 
the flavor SU(3) symmetry breaking and those in the SU(3) symmetric
case, respectively.}  
\label{fig:2}
\end{figure}
As shown in Eqs.~\eqref{eq:relations1} and~\eqref{eq:relations2}, we
have only two independent axial-vector transition form factors when
the flavor SU(3) symmetry is considered. We will present the results
for these two form factors. 
In left panel of Fig.~\ref{fig:2}, we draw the results for the
axial-vector transition form factors for the $\Sigma_c^+\to
\Lambda_c^+$ transition. All other form factors for the
axial-vector transitions $B_{1/2}\to B_{1/2}'$ are related to that for
the $\Sigma_c^+\to \Lambda_c^+$ transition. So, we take the 
$\Sigma_c^+\to \Lambda_c^+$ transition form factor as a prototype
one. The dashed curve represents that in the flavor SU(3) symmetric
case, while the solid one depicts that with linear $m_{\mathrm{s}}$
corrections. The effects of SU(3) symmetry breaking are tiny. This can
be understood by examining
Eqs.~\eqref{eq:GA3sym}--\eqref{eq:GA3wfcorr}. The prefactors in
Eqs.~\eqref{eq:GA3opcorr} and~\eqref{eq:GA3wfcorr} are much smaller
than that in the leading-order contribution given in
Eq.~\eqref{eq:GA3sym}. In addition, the first term in the bracket of
Eq.~\eqref{eq:GA3sym} is the most dominant one.  In the right panel of
Fig.~\ref{fig:2}, we depict the results for the axial-vector
transition form factors from $\Sigma_c^{*++}$ belonging to the baryon
sextet with spin 3/2 to $\Sigma_c^{++}$ in the sextet with spin
1/2. Note that the expression for the leading-order contribution to
$C_{5,B\to B'}^{A}(Q^2)$ is distinguished from that for $G_{A,B\to
  B'}(Q^2)$ by the last term in Eq.~\eqref{eq:C53sym}, which is
proportional to $\mathcal{B}_0^{B\to B'}$ and $\mathcal{B}_2^{B\to B'}$.   
They provide about $9~\%$ corrections to $C_{5,B\to B'}^{A}(Q^2)$. The
effects of the flavor SU(3) symmetry breaking turn out very
small. Thus, we will neglect them in the discussion of other
observables related to the axial-vector transition form factors. 

\begin{figure}[htp]
  \includegraphics[scale=0.52]{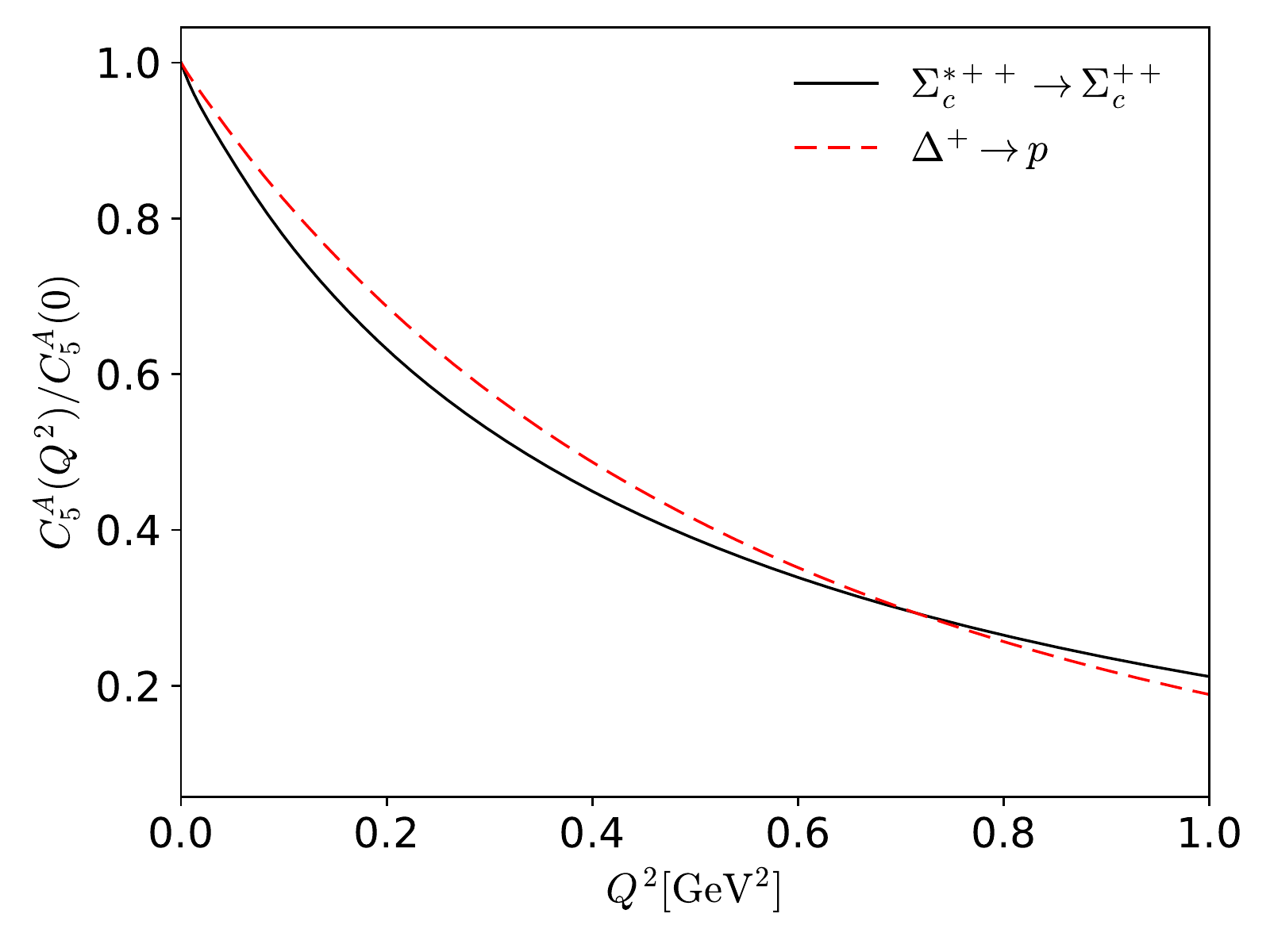}
\caption{Comparison of $Q^{2}$ dependence of $C_{5, \; \Sigma_c^{*++}
    \rightarrow \Sigma_c^{++}}^{A}(Q^{2})$ to $C_{5, \; \Delta^{+}
    \rightarrow p}^{A}(Q^{2})$. The solid curve represents the result
  for $C_{5, \; \Sigma_c^{*++} \rightarrow \Sigma_c^{++}}^{A}(Q^{2})$,
  whereas the dashed one depicts $C_{5, \; \Delta^{+}
    \rightarrow p}^{A}(Q^{2})$.} 
\label{fig:3}
\end{figure}
It is of great interest to compare the $Q^2$ dependence of the result
for $C_{5, \; \Sigma_c^{*++} \rightarrow
\Sigma_c^{++}}^{A}(Q^{2})$ to that for $C_{5, \; \Delta^{+}
  \rightarrow p}^{A}(Q^{2})$, since both form factors describe the
axial-vector transitions from the spin-3/2 baryon to the spin-1/2
baryon. To compare more closely, we normalize them by the
corresponding values of the form factors at $Q^2=0$. As shown in
Fig.~\ref{fig:3}, the axial-vector form factor for the
$\Sigma_c^{*++} \rightarrow \Sigma_c^{++}$ transition starts to fall
off more fast than that for the $\Delta^+\to p$ transition. It
indicates that the mean square radius for the $\Sigma_c^{*++}
\rightarrow \Sigma_c^{++}$ transition is larger than that for the
$\Delta^+\to p$ one. We want to emphasize that the pion mean field for
the singly heavy baryons is different from that for the light
baryons. This makes main difference between the $\Sigma_c^{*++}
\rightarrow \Sigma_c^{++}$ and $\Delta^+\to p$ transition form factors,
as exhibited in Fig.~\ref{fig:3}.  

\begin{table}[htp]
 \renewcommand{\arraystretch}{1.4}
 {\setlength{\tabcolsep}{13pt}
 \caption{Numerical results for the axial-vector transition constants
  $G_{A,\,\Sigma_c^+ \to \Lambda_{c}^+}(0)(C^{A}_{5,\,\Sigma_c^{*++}
    \to \Sigma_c^{++}}(0))$, and the corresponding axial masses 
  and mean square radii.}   
 \label{tab:1}
 \begin{tabularx}{0.65\linewidth}{ c | c c} 
  \hline 
  \hline 

 & $\Sigma_{c}^{+} \rightarrow \Lambda_{c}^{+}$
& $\Sigma_{c}^{*++} \rightarrow \Sigma_{c}^{++}$
\\
  \hline 
$G_{A,\,\Sigma_c^+ \to \Lambda_{c}^+}(0)(C^{A}_{5,\,\Sigma_c^{*++}
    \to \Sigma_c^{++}}(0))$  
& $-0.955$   & $-0.443$ \\ 
\hline 
$M_{A}$ [GeV] 
& $0.759$   & $0.728$\\
$\langle r^{2}_{A}\rangle$ [fm$^{2}$] 
& $0.812$   & $0.882$\\
\hline  
 \hline
\end{tabularx}}
\end{table}
In Table~\ref{tab:1}, we list the values of the $G_{A,\,\Sigma_c^+ \to
  \Lambda_{c}^+}(0)$ and $C^{A}_{5,\,\Sigma_c^{*++} \to
  \Sigma_c^{++}}(0)$ at $Q^2=0$. These values will be used for
determining the decay widths for the strong decays of the singly heavy
baryons. The axial-vector transition form factor $G_{A,\,\Sigma_c^+ \to
  \lambda_{c}^+}(Q^2)$ presented in
Fig.~\ref{fig:2} can be parametrized by the dipole-type
parametrization 
\begin{align}
G_{A} = \frac{G_A(0)}{(1+Q^2/M_A^2)^2},
\end{align}
where $M_A$ is called the axial mass. We can parametrize
$C^{A}_{5,\,\Sigma_c^{*++} \to  \Sigma_c^{++}}(Q^2)$ in the same
manner. The numerical results for $M_A$ are given in the third row in
Table~\ref{tab:1}, which indicates that the $Q^2$ dependence of the
$\Sigma_c^+ \to \Lambda_{c}^+$ and $\Sigma_c^{*++} \to
  \Sigma_c^{++}$ form factors are similar. The results for the mean
  square radii are listed in the last row of Table~\ref{tab:1}. 

\begin{table}[htp]
\centering
 \renewcommand{\arraystretch}{1.4}
 \setlength{\tabcolsep}{14pt}
 \caption{Numerical results for the strong decay widths in
  comparison with the experimental data.}  
   \label{tab:2}
   {\footnotesize
 \begin{tabularx}{1.0\linewidth}{ c | c | c c c c c c c} 
  \hline 
  \hline 
Decay modes &
 $\Gamma$ [MeV] 
 & Exp.~\cite{PDG} & FOCUS Coll.~\cite{FOCUS:2001qdu} & CLEO II~\cite{CLEO:2001zwr} 
 &   Belle~\cite{Belle:2013htj,Belle:2014fde,Belle:2021qip}  \\
  \hline 
$\Sigma_{c}^{++} \rightarrow \Lambda_{c}^{+}+\pi^{+}$ 
& $2.80$ 
& $1.89^{+0.09}_{-0.18}$ & $2.05 ^{+0.41}_{-0.38}$ & $2.3 \pm 0.2 \pm 0.3$ &$1.84 \pm 0.04^{+0.07}_{-0.20}$  \\
$\Sigma_{c}^{+} \rightarrow \Lambda_{c}^{+}+\pi^{0}$ 
& $3.39$ 
&  $< 4.6$ & - & - & $2.3 \pm 0.3 \pm 0.3$  \\
$\Sigma_{c}^{0} \rightarrow \Lambda_{c}^{+}+\pi^{-}$ 
& $2.76$ 
& $1.83^{+0.11}_{-0.19}$ & $1.55^{+0.41}_{-0.37}$ & $2.5 \pm 0.2 \pm 0.3$ & $1.76 \pm 0.04 ^{+0.09}_{-0.21}$ \\
$\Sigma_{c}^{*++} \rightarrow \Lambda_{c}^{+}+\pi^{+}$ 
& $21.0$ 
& $14.78^{+0.30}_{-0.40}$ & - & - & $14.77 \pm 0.25^{+0.18}_{-0.30}$ \\
$\Sigma_{c}^{*+} \rightarrow \Lambda_{c}^{+}+\pi^{0}$ 
& $22.1$ 
& $< 17$ & - & - & $17.2 ^{+2.3+3.1}_{-2.1-0.7}$   \\
$\Sigma_{c}^{*0} \rightarrow \Lambda_{c}^{+}+\pi^{-}$ 
& $21.0$ 
& $15.3^{+0.4}_{-0.5}$ & - & - & $15.41 \pm 0.41 ^{+0.20}_{-0.32}$ \\
$\Xi_{c}^{*+} \rightarrow \Xi_{c}+\pi$ 
& $2.12$ 
 & $2.14 \pm 0.19$ & - & - & $2.6 \pm 0.2 \pm 0.4$ \\
$\Xi_{c}^{*0} \rightarrow \Xi_{c}+\pi$ 
& $2.30$ 
& $2.35 \pm 0.22$ & - &  - & - \\
\hline
\hline
\end{tabularx}
}
\end{table}
\begin{table}[htp]
\centering
 \renewcommand{\arraystretch}{1.4}
 \setlength{\tabcolsep}{2.7pt}
 \caption{Numerical results for the strong decay widths in
  comparison with those from various works.}  
   \label{tab:3}
   {\footnotesize
 \begin{tabularx}{1.0\linewidth}{ c | c |c c c c c c c} 
  \hline 
  \hline 
Decay modes &
$\Gamma$ [MeV]  & Yan et al.~\cite{Yan:1992gz}&Huang et al.~\cite{Huang:1995ke} &Rosner~\cite{Rosner:1995yu} 
 & Pirjol et al.~\cite{Pirjol:1997nh} & Tawfiq et al.~\cite{Tawfiq:1998nk}& Ivanov et al.~\cite{Ivanov:1999bk}       \\ 
\hline
$\Sigma_{c}^{++} \rightarrow \Lambda_{c}^{+}+\pi^{+}$ 
& $2.80$ 
& - & $2.5$ & $1.32 \pm 0.04$ 
& $2.025^{+1.134}_{-0.987}$ & $1.64$ & $2.85 \pm 0.19$\\ 
$\Sigma_{c}^{+} \rightarrow \Lambda_{c}^{+}+\pi^{0}$
& $3.39$ 
&- & $3.2$ & $1.32 \pm 0.04$ 
& - & $1.70$ & $3.63 \pm 0.27$\\ 
$\Sigma_{c}^{0} \rightarrow \Lambda_{c}^{+}+\pi^{-}$
& $2.76$ 
& $2.45, \, 4.35$ & $2.4$ & $1.32 \pm 0.04$ 
& $1.939^{+1.114}_{-0.954}$ & $1.57$ & $2.65 \pm 0.19$ \\ 
$\Sigma_{c}^{*++} \rightarrow \Lambda_{c}^{+}+\pi^{+}$
& $21.0$ 
& - & $25$ & $20$ 
& - & $12.84$ & $21.99 \pm 0.87$\\ 
$\Sigma_{c}^{*+} \rightarrow \Lambda_{c}^{+}+\pi^{0}$ 
& $22.1$ 
& - & $25$ & $20$ 
& - & - & - \\ 
$\Sigma_{c}^{*0} \rightarrow \Lambda_{c}^{+}+\pi^{-}$
& $21.0$ 
& - & $25$ & $20$ 
& - & $12.40$ & $21.21 \pm 0.81$\\ 
$\Xi_{c}^{*+} \rightarrow \Xi_{c}+\pi$ 
& $2.12$ 
& - & - & $2.3 \pm 0.1$ 
& - & $1.12$ & $1.78 \pm 0.33$\\ 
$\Xi_{c}^{*0} \rightarrow \Xi_{c}+\pi$
& $2.30$ 
& - & -  & $2.3 \pm 0.1$ 
& - & $1.16$ & $2.11 \pm 0.29$\\ 
\hline
\hline
\end{tabularx}
}
\end{table}
\begin{table}[htp]
\centering
 \renewcommand{\arraystretch}{1.4}
 \setlength{\tabcolsep}{1.0pt}
 \caption{Numerical results for the strong decay widths in
  comparison with those from various works.}
   \label{tab:4}
 {\footnotesize
 \begin{tabularx}{1.0\linewidth}{ c |c | c c c c c c c} 
  \hline 
  \hline 
Decay modes & 
$\Gamma$ [MeV] & Albertus et al.\cite{Albertus:2005zy} & Chen et al.\cite{Chen:2007xf} &Azizi et al.\cite{Azizi:2008ui} 
  & Cheng et al.\cite{Cheng:2015naa}& Nagahiro et al.\cite{Nagahiro:2016nsx} & Can et al.\cite{Can:2016ksz}\\ 
\hline
$\Sigma_{c}^{++} \rightarrow \Lambda_{c}^{+}+\pi^{+}$ 
& $2.80$ 
& $2.41 \pm 0.07 \pm 0.02$ & $1.24$ & $2.16 \pm 0.85$
& - & $4.27 - 4.33$ & $1.65 \pm 0.28 \pm 0.30$\\ 
$\Sigma_{c}^{+} \rightarrow \Lambda_{c}^{+}+\pi^{0}$
& $3.39$ 
& $2.79 \pm 0.08 \pm 0.02$ & $1.40$ & $2.16 \pm 0.85$
& $2.3^{+0.1}_{-0.2}$ &- & $1.65 \pm 0.28 \pm 0.30$\\ 
$\Sigma_{c}^{0} \rightarrow \Lambda_{c}^{+}+\pi^{-}$
& $2.76$ 
& $2.37 \pm 0.07 \pm 0.02$ & $1.24$ & $2.16 \pm 0.85$
& $1.9^{+0.1}_{-0.2}$ & - & $1.65 \pm 0.28 \pm 0.30$\\ 
$\Sigma_{c}^{*++} \rightarrow \Lambda_{c}^{+}+\pi^{+}$
& $21.0$ 
& $17.52 \pm 0.74 \pm 0.12$ & $11.9$ & - 
& $14.5^{+0.5}_{-0.8}$ & $30.3 - 31.6$ & - \\ 
$\Sigma_{c}^{*+} \rightarrow \Lambda_{c}^{+}+\pi^{0}$
& $22.1$ 
 & $17.31 \pm 0.73 \pm 0.12$ & $12.1$ & -
& $15.2^{+0.6}_{-1.3}$ & - & - \\ 
$\Sigma_{c}^{*0} \rightarrow \Lambda_{c}^{+}+\pi^{-}$
& $21.0$  
& $16.90 \pm 0.71 \pm 0.12$ & $11.9$ & -
& $14.7^{+0.6}_{-1.2}$ & - & - \\ 
$\Xi_{c}^{*+} \rightarrow \Xi_{c}+\pi$ 
& $2.12$ 
& $1.84 \pm 0.06 \pm 0.01$ & $0.64$ & -
& $2.4^{+0.1}_{-0.2}$& - & - \\ 
$\Xi_{c}^{*0} \rightarrow \Xi_{c}+\pi$
& $2.30$ 
& $2.07 \pm 0.07 \pm 0.01$ & $0.54$ & -
& $2.5^{+0.1}_{-0.2}$& - & - \\ 
\hline
\hline
\end{tabularx}
}
\end{table}
In Table~\ref{tab:2}, we compare the current results for the strong
decay widths of the singly heavy baryons, of which the experimental
data are available. Thus, we consider the strong decays of $\Sigma_c$,
$\Sigma_c^*$, and $\Xi_c^*$. The third column of Table~\ref{tab:2} are
the experimental data taken from the PDG~\cite{PDG}. In the second,
third, and fourth columns of Table~\ref{tab:2}, we list the
experimental data taken from the FOCUS
Collaboration~\cite{FOCUS:2001qdu}, the CLEO
Collaboration~\cite{CLEO:2001zwr}, and the Belle
Collaboration~\cite{Belle:2013htj,Belle:2014fde,Belle:2021qip},
respectively. The values of the decay widths for the strong decay of
$\Sigma_c$ and $\Sigma_c^*$ are overestimated, compared with the
experimental data. On the other hand, those of the $\Xi_c^*$ decays
are in good agreement with the data. In Tables~\ref{tab:3}
and~\ref{tab:4}, we compare the current results with those from other
works. We find that the results from the present work are consistent
with those from Refs.~\cite{Ivanov:1999bk, Albertus:2005zy}.

\section{Summary and conclusion} 
\label{sec:6}
We investigated the axial-vector transition form factors of the baryon
sextet within the framework of the chiral quark-soliton
model. Assuming that the heavy-quark mass is infinitely heavy, the
$N_c-1$ light valence quarks govern the quark dynamics inside a singly
heavy baryon. In contrast, the singly heavy quark is merely
a static color source, making singly heavy baryons
the color singlet. The presence of the $N_c-1$ valence quarks creates
the pion mean field, so that they are also influenced by it
self-consistently. The $N_c-1$ valence quarks also constrain the right
hypercharge $Y_R=(N_c-1)/3=2/3$, allowing the flavor SU(3)
representations such as the baryon antitriplet, baryon sextet, and
higher representations. Based on this framework, we studied the  
axial-vector transitions from the baryon sextet with spin 1/2 and 3/2
to the baryon sextet with spin 1/2 and antitriplet, considering the
rotational $1/N_c$ and linear $m_{\mathrm{s}}$ corrections.  
We presented the results for the $\Sigma_c^+\to \Lambda_c^+$ and
$\Sigma_c^{*++}\to \Sigma_c^{++}$ form factors. These are the first
results for the axial-vector transition form factors of the singly
heavy baryons. Those for all other decay channels are related either
by isospin symmetry or by flavor SU(3) symmetry. 
We found that the effects of the flavor SU(3) symmetry breaking were
tiny. Thus, we neglected them to compute other observables.
We also obtained the corresponding axial masses employing the
dipole-type parametrizations for the form factors. We derived the
axial-transition mean square radii.    
Using the values of the form factors at $Q^2=0$, we got the decay
rates for the strong decays of $\Sigma_c$, $\Sigma_c^*$, and
$\Xi_c^*$. The decay rates of $\Sigma_c$ and $\Sigma_c^*$
decays are overestimated in comparison with the data but those of the
$\Xi_c^*$ decays are in good agreement with the data. 

\begin{acknowledgments}
The authors are grateful to Gh.-S. Yang and J.-Y. Kim for fruitful
discussions. We also want to express our gratitude to Y.-S. Jun for
his contribution to the initial stage of the current work. 
The present work was supported by an Inha University
Research Grant in 2022.
\end{acknowledgments} 
\clearpage
\appendix
\section{Components of the axial-vector transition form
  factors} \label{app:a} 
In this Appendix, the expressions for the axial-vector transition form
factors in Eqs.~\eqref{eq:GAform},~\eqref{eq:C5form} will be given 
explicitly 
\begin{align}
\mathcal{A}^{B \rightarrow B'}_{0}(Q^{2}) &= 
  \frac{\sqrt{M_{B'}}}{\sqrt{E_{B'}+M_{B'}}} 
  \int d^{3} r j_{0}(|\bm{q}| |\bm{r}|) \left[(N_{c}-1) \phi^{\dagger}_{\mathrm{val}}
  (\bm{r}) \bm{\sigma} \cdot \bm{\tau} \phi_{\mathrm{val}}(r)\right. \cr
& \left. \hspace{6cm} +\; N_{c} \sum_{n} \phi^{\dagger}_{n}(\bm{r})
  \bm{\sigma} \cdot \bm{\tau}  
  \phi_{n}(\bm{r}) \mathcal{R}_{1}(E_n) \right] ,\\
\mathcal{B}^{B \rightarrow B'}_{0}(Q^{2}) &= 
  \frac{\sqrt{M_{B'}}}{\sqrt{E_{B'}+M_{B'}}} 
  \int d^{3} r j_{0}(|\bm{q}| |\bm{r}|) \left[(N_{c}-1) \sum_{n \ne
  \mathrm{val} } \frac{1}{E_{\mathrm{val}}-E_{n}} 
  \phi^{\dagger}_{\mathrm{val}}(\bm{r}) \bm{\sigma} 
  \phi_{n}(\bm{r}) \cdot \langle n | \bm{\tau} | \mathrm{val} \rangle 
  \right. \cr
& \left. \hspace{3.9cm} 
  -\frac{1}{2}N_{c} \sum_{n,m} \phi^{\dagger}_{n}(\bm{r}) \bm{\sigma} 
  \phi_{m}(\bm{r}) \cdot \langle m | \bm{\tau} | n \rangle 
  \mathcal{R}_{5}(E_n,E_m) \right], \\
\mathcal{C}^{B \rightarrow B'}_{0}(Q^{2}) &= 
  \frac{\sqrt{M_{B'}}}{\sqrt{E_{B'}+M_{B'}}} 
  \int d^{3} r j_{0}(|\bm{q}| |\bm{r}|) \left[(N_{c}-1) \sum_{n_{0} \ne
  \mathrm{val} } \frac{1}{E_{\mathrm{val}}-E_{n_{0}}} 
  \phi^{\dagger}_{\mathrm{val}}(\bm{r}) \bm{\sigma} \cdot \bm{\tau} 
  \phi_{n_{0}}(\bm{r}) \langle n_{0} | \mathrm{val} \rangle \right. \cr
& \left. \hspace{3.9cm} 
  -N_{c}\sum_{n,m_{0}} \phi^{\dagger}_{n}(\bm{r}) \bm{\sigma} \cdot \bm{\tau} 
  \phi_{m_{0}}(\bm{r}) \langle m_{0} | n \rangle 
  \mathcal{R}_{5}(E_n,E_{m_{0}}) \right], \\
\mathcal{D}^{B \rightarrow B'}_{0}(Q^{2}) &= 
  \frac{\sqrt{M_{B'}}}{\sqrt{E_{B'}+M_{B'}}} 
  \int d^{3} r j_{0}(|\bm{q}| |\bm{r}|) \left[(N_{c}-1) \sum_{n \ne
  \mathrm{val} } \frac{\mathrm{sgn}(E_{n})}{E_{\mathrm{val}}-E_{n}} 
  \phi^{\dagger}_{\mathrm{val}}(\bm{r}) (\bm{\sigma} \times \bm{\tau})
  \phi_{n}(\bm{r}) \cdot \langle n | \bm{\tau} | \mathrm{val} 
  \rangle \right. \cr
& \left. \hspace{3.9cm} 
  + \frac{1}{2}N_{c} \sum_{n,m} \phi^{\dagger}_{n}(\bm{r}) \bm{\sigma} 
  \times \bm{\tau} \phi_{m}(\bm{r}) \cdot \langle m | \bm{\tau} 
  | n \rangle \mathcal{R}_{4}(E_n,E_m) \right],\\
\mathcal{H}^{B \rightarrow B'}_{0}(Q^{2}) &= 
  \frac{\sqrt{M_{B'}}}{\sqrt{E_{B'}+M_{B'}}} 
  \int d^{3} r j_{0}(|\bm{q}| |\bm{r}|) \left[(N_{c}-1) \sum_{n \ne
  \mathrm{val} } \frac{1}{E_{\mathrm{val}}-E_{n}} 
  \phi^{\dagger}_{\mathrm{val}}(\bm{r}) \bm{\sigma} \cdot \bm{\tau} 
  \langle \bm{r} | n \rangle \langle n | \gamma^{0}| \mathrm{val} 
  \rangle \right. \cr
& \left. \hspace{3.9cm} 
  + \frac{1}{2}N_{c} \sum_{n,m} \phi^{\dagger}_{n}(\bm{r}) \bm{\sigma} \cdot 
  \bm{\tau} \phi_{m}(\bm{r}) \langle m | \gamma^{0} | n \rangle 
  \mathcal{R}_{2}(E_n,E_m) \right], \\
\mathcal{I}^{B \rightarrow B'}_{0}(Q^{2}) &= 
  \frac{\sqrt{M_{B'}}}{\sqrt{E_{B'}+M_{B'}}} 
  \int d^{3} r j_{0}(|\bm{q}| |\bm{r}|) \left[(N_{c}-1) \sum_{n \ne
  \mathrm{val} } \frac{1}{E_{\mathrm{val}}-E_{n}} 
  \phi^{\dagger}_{\mathrm{val}}(\bm{r}) \bm{\sigma} \phi_{n}(\bm{r}) 
  \cdot \langle n | \gamma^{0} \bm{\tau} | \mathrm{val} \rangle 
  \right. \cr
& \left. \hspace{3.9cm} +\frac{1}{2}N_{c} \sum_{n,m} \phi^{\dagger}_{n}
  (\bm{r}) \bm{\sigma} \phi_{m}(\bm{r}) \cdot \langle m | \gamma^{0} 
  \bm{\tau} | n \rangle \mathcal{R}_{2}(E_n,E_m)\right],\\
\mathcal{J}^{B \rightarrow B'}_{0}(Q^{2}) &= 
  \frac{\sqrt{M_{B'}}}{\sqrt{E_{B'}+M_{B'}}} 
  \int d^{3} r j_{0}(|\bm{q}| |\bm{r}|) \left[ (N_{c}-1)\sum_{n_{0} \ne
  \mathrm{val} } \frac{1}{E_{\mathrm{val}}-E_{n_{0}}} 
  \phi^{\dagger}_{\mathrm{val}}(\bm{r}) \bm{\sigma} \cdot \bm{\tau} 
  \phi_{n_{0}}(\bm{r}) \langle n_{0}| \gamma^{0} | \mathrm{val} \rangle 
  \right. \cr
& \left. \hspace{3.9cm} +N_{c} \sum_{n,m_{0}} \phi^{\dagger}_{n}(\bm{r}) 
  \bm{\sigma} \cdot \bm{\tau} \phi_{m_{0}}(\bm{r}) 
  \langle m_{0}| \gamma^{0} | n \rangle \mathcal{R}_{2}(E_n,E_{m_{0}}) 
  \right],
\label{AxComp10} \\
\mathcal{A}^{B \rightarrow B'}_{2}(Q^{2}) &= 
  \frac{\sqrt{M_{B'}}}{\sqrt{E_{B'}+M_{B'}}} 
  \int d^{3} r j_{2}(|\bm{q}| |\bm{r}|) \left[(N_{c}-1) \phi^{\dagger}_{\mathrm{val}}
  (\bm{r}) \left\{ \sqrt{2\pi}Y_{2} \otimes \sigma_{1}\right\}_{1} \cdot 
  \bm{\tau} \phi_{\mathrm{val}}(\bm{r}) \right. \cr
& \left. \hspace{3.9cm} 
  +N_{c}\sum_{n} \phi^{\dagger}_{n}(\bm{r}) \left\{ \sqrt{2\pi}Y_{2} 
  \otimes \sigma_{1}\right\}_{1} \cdot \bm{\tau}
  \phi_{n}(\bm{r}) \mathcal{R}_{1}(E_n) \right] ,
  \end{align}
  \begin{align}
\mathcal{B}^{B \rightarrow B'}_{2}(Q^{2}) &= 
  \frac{\sqrt{M_{B'}}}{\sqrt{E_{B'}+M_{B'}}} 
  \int d^{3} r j_{2}(|\bm{q}| |\bm{r}|) \left[(N_{c}-1) \sum_{n \ne
  \mathrm{val} } \frac{\langle n | \bm{\tau} | \mathrm{val}
    \rangle}{E_{\mathrm{val}}-E_{n}}  \cdot
  \phi^{\dagger}_{\mathrm{val}}(\bm{r}) \left\{ \sqrt{2\pi}Y_{2} \otimes 
  \sigma_{1}\right\}_{1} \phi_{n}(\bm{r})
   \right. \cr
& \left. \hspace{3.9cm} 
  -\frac{1}{2}N_{c} \sum_{n,m} \phi^{\dagger}_{n}(\bm{r}) 
  \left\{ \sqrt{2\pi}Y_{2} \otimes \sigma_{1}\right\}_{1} 
  \phi_{m}(\bm{r}) \cdot \langle m | \bm{\tau} | n \rangle 
  \mathcal{R}_{5}(E_n,E_m) \right], \\
\mathcal{C}^{B \rightarrow B'}_{2}(Q^{2}) &= 
  \frac{\sqrt{M_{B'}}}{\sqrt{E_{B'}+M_{B'}}} 
  \int d^{3} r j_{2}(|\bm{q}| |\bm{r}|) \left[(N_{c}-1) \sum_{n_{0} \ne
  \mathrm{val} } \frac{ \langle n_{0} |
   \mathrm{val} \rangle}{E_{\mathrm{val}}-E_{n_{0}}}  
  \phi^{\dagger}_{\mathrm{val}}(\bm{r}) \left\{ \sqrt{2\pi}Y_{2} \otimes 
  \sigma_{1}\right\}_{1} \cdot \bm{\tau} \phi_{n_{0}}(\bm{r}) 
  \right. \cr
& \left. \hspace{3.9cm} 
  -N_{c}\sum_{n,m_{0}} \phi^{\dagger}_{n}(\bm{r}) \left\{ \sqrt{2\pi}Y_{2} 
  \otimes \sigma_{1}\right\}_{1} \cdot \bm{\tau} 
  \phi_{m_{0}}(\bm{r}) \langle m_{0} | n \rangle 
  \mathcal{R}_{5}(E_n,E_{m_{0}}) \right], \\
\mathcal{D}^{B \rightarrow B'}_{2}(Q^{2}) &= 
  \frac{\sqrt{M_{B'}}}{\sqrt{E_{B'}+M_{B'}}} 
  \int d^{3} r j_{2}(|\bm{q}| |\bm{r}|) \left[(N_{c}-1) \sum_{n \ne
  \mathrm{val} } \frac{\mathrm{sgn}(E_{n})\langle n | \bm{\tau} |
     \mathrm{val} 
  \rangle}{E_{\mathrm{val}}-E_{n}}   \cdot
  \phi^{\dagger}_{\mathrm{val}}(\bm{r}) \left\{ \sqrt{2\pi}Y_{2} \otimes 
  \sigma_{1}\right\}_{1} \times \bm{\tau} 
  \phi_{n}(\bm{r})   \right. \cr
& \left. \hspace{3.9cm} 
  + \frac{1}{2}N_{c} \sum_{n,m} \phi^{\dagger}_{n}(\bm{r}) 
  \left\{ \sqrt{2\pi}Y_{2} \otimes \sigma_{1}\right\}_{1} 
  \times \bm{\tau} \phi_{m}(\bm{r}) \cdot \langle m | \bm{\tau} 
  | n \rangle \mathcal{R}_{4}(E_n,E_m) \right],\\
\mathcal{H}^{B \rightarrow B'}_{2}(Q^{2}) &= 
  \frac{\sqrt{M_{B'}}}{\sqrt{E_{B'}+M_{B'}}} 
  \int d^{3} r j_{2}(|\bm{q}| |\bm{r}|) \left[(N_{c}-1) \sum_{n \ne
  \mathrm{val} } \frac{ \langle n | \gamma^{0}| \mathrm{val}
    \rangle}{E_{\mathrm{val}}-E_{n}} 
  \phi^{\dagger}_{\mathrm{val}}(\bm{r}) \left\{ \sqrt{2\pi}Y_{2} \otimes 
  \sigma_{1}\right\}_{1} \cdot \bm{\tau} \phi_{n}(\bm{r})
 \right. \cr
& \left. \hspace{3.9cm} 
  + \frac{1}{2}N_{c} \sum_{n,m} \phi^{\dagger}_{n}(\bm{r}) 
  \left\{ \sqrt{2\pi}Y_{2} \otimes \sigma_{1}\right\}_{1} \cdot 
  \bm{\tau} \phi_{m}(\bm{r}) \langle m | \gamma^{0} | n \rangle 
  \mathcal{R}_{2}(E_n,E_m) \right], \cr\\
\mathcal{I}^{B \rightarrow B'}_{2}(Q^{2}) &= 
  \frac{\sqrt{M_{B'}}}{\sqrt{E_{B'}+M_{B'}}} 
  \int d^{3} r j_{2}(|\bm{q}| |\bm{r}|) \left[(N_{c}-1) \sum_{n \ne
  \mathrm{val} } \frac{ \langle n | \gamma^{0} \bm{\tau} |
     \mathrm{val}  \rangle}{E_{\mathrm{val}}-E_{n}}  \cdot 
  \phi^{\dagger}_{\mathrm{val}}(\bm{r}) \left\{ \sqrt{2\pi}Y_{2} \otimes 
  \sigma_{1}\right\}_{1} \phi_{n}(\bm{r}) 
  \right. \cr
& \left. \hspace{3.9cm} +\frac{1}{2} N_{c}\sum_{n,m} 
  \phi^{\dagger}_{n}(\bm{r}) \left\{ \sqrt{2\pi}Y_{2} \otimes 
  \sigma_{1}\right\}_{1} \phi_{m}(\bm{r}) \cdot \langle m | \gamma^{0} 
  \bm{\tau} | n \rangle \mathcal{R}_{2}(E_n,E_m)\right], \\
\mathcal{J}^{B \rightarrow B'}_{2}(Q^{2}) &= 
  \frac{\sqrt{M_{B'}}}{\sqrt{E_{B'}+M_{B'}}}
  \int d^{3} r j_{2}(|\bm{q}| |\bm{r}|) \left[(N_{c}-1) \sum_{n_{0} \ne
  \mathrm{val} } \frac{\langle n_{0}| \gamma^{0} | \mathrm{val}
     \rangle  }{E_{\mathrm{val}}-E_{n_{0}}}  
  \phi^{\dagger}_{\mathrm{val}}(\bm{r}) 
  \left\{ \sqrt{2\pi}Y_{2} \otimes \sigma_{1}\right\}_{1} \cdot \bm{\tau} 
  \phi_{n_{0}}(\bm{r}) 
  \right. \cr
& \left. \hspace{3.9cm}
  +N_{c} \sum_{n,m_{0}} \phi^{\dagger}_{n}(\bm{r}) 
  \left\{ \sqrt{2\pi}Y_{2} \otimes \sigma_{1}\right\}_{1} \cdot \bm{\tau} 
  \phi_{m_{0}}(\bm{r}) \langle m_{0}| \gamma^{0} | n \rangle 
  \mathcal{R}_{2}(E_n,E_{m_{0}}) \right],
\label{AxComp12}
\end{align}
where the regularization functions are defined by 
\begin{align}
\mathcal{R}_{1}(E_{n}) &= \frac{-E_{n}}{2 \sqrt{\pi}} \int^{\infty}_{0}
  \phi(u) \frac{du}{\sqrt{u}} e^{-u E_{n}^{2}}, \\
\mathcal{R}_{2}(E_{n},E_{m}) &= \frac{1}{2 \sqrt{\pi}} \int^{\infty}_{0}
  \phi(u) \frac{du}{\sqrt{u}} \frac{ E_{m} e^{-u E_{m}^{2}} 
  -E_{n}e^{-uE_{n}^{2}}}{E_{n} - E_{m}}, \\
\mathcal{R}_{4}(E_{n},E_{m}) &= \frac{1}{2 \pi} \int^{\infty}_{0} du
  \, \phi(u) \int^{1}_{0} d\alpha e^{-\alpha uE^{2}_{m} 
  -(1-\alpha)uE^{2}_{n}} \frac{(1-\alpha)E_{n}-\alpha E_{m}}
  {\sqrt{\alpha(1-\alpha)}}, \\
\mathcal{R}_{5}(E_{n},E_{m}) &=
  \frac{\mathrm{sgn}(E_{n})-\mathrm{sgn}(E_{m})}{2(E_{n}-E_{m})}.
\end{align}
Here, $|\mathrm{val}\rangle$ and $|n\rangle$ represent the state of 
the valence and sea quarks with the corresponding eigenenergies
$E_{\mathrm{val}}$ and $E_n$ of the one-body Dirac 
Hamiltonian $h(U)$, respectively.

\section{Matrix elements of the SU(3) Wigner ${D}$
  function}\label{app:b} 
In the following we list the results for the matrix elements of the 
relevant collective operators for the axial-vector transition form
factors of the singly heavy baryons in Table~\ref{tab:5} 
to \ref{tab:12}. 
\begin{table}[htp]
\setlength{\tabcolsep}{5pt}
\renewcommand{\arraystretch}{2.2}
  \caption{The matrix elements of the single and double
  Wigner $D$ function operators when $a=3$.}
  \label{tab:5}
\begin{center}
\begin{tabular}{ c | c c c c c c c}
 \hline
  \hline
$B \rightarrow B'$
  & $\Sigma_{c}^{+} \rightarrow \Lambda_{c}^{+}$ & $\Xi_{c}^{\prime}  \rightarrow \Xi_{c}$ &
  $\Sigma_{c}^{*+} \rightarrow \Lambda_{c}^{+}$ & $\Xi_{c}^{*}  \rightarrow \Xi_{c}$ &
  $\Sigma_{c}^{*} \rightarrow \Sigma_{c}$ & $\Xi_{c}^{*} \rightarrow \Xi_{c}^{\prime}$ &
  $\Omega_{c}^{*0} \rightarrow \Omega_{c}^{0}$\\
 \hline
$\langle B' |D^{(8)}_{33} | B \rangle$
  & $-\frac{1}{2\sqrt{6}}$ & $\frac{1}{2\sqrt{6}} T_{3}$
  & $\frac{1}{2\sqrt{3}}$ & $-\frac{1}{2\sqrt{3}} T_{3}$ 
  & $-\frac{1}{5\sqrt{2}} T_{3}$ & $-\frac{1}{5\sqrt{2}} T_{3}$
  & $0$ \\
$\langle B' |D^{(8)}_{38} \hat{J}_{3} | B \rangle$
  & $0$ & $0$
  & $0$ & $0$
  & $\frac{1}{5\sqrt{6}} T_{3}$ & $\frac{1}{5\sqrt{6}} T_{3}$ 
  & $0$ \\
$\langle B' |d_{bc3} D^{(8)}_{3b} \hat{J}_{c} | B \rangle$
  & $\frac{1}{4\sqrt{6}}$ & $-\frac{1}{4\sqrt{6}} T_{3}$
  & $-\frac{1}{4\sqrt{3}}$ & $\frac{1}{4\sqrt{3}} T_{3}$ 
  & $\frac{1}{10\sqrt{2}} T_{3}$ & $\frac{1}{10\sqrt{2}} T_{3}$ 
  & $0$ \\
$\langle B' |D^{(8)}_{83}D^{(8)}_{38} | B \rangle$
  & $-\frac{1}{20\sqrt{6}}$ & $\frac{1}{5\sqrt{6}} T_{3}$
  & $\frac{1}{20\sqrt{3}}$ & $-\frac{1}{5\sqrt{3}} T_{3}$
  & $-\frac{\sqrt{2}}{45} T_{3}$ & $-\frac{1}{45\sqrt{2}} T_{3}$
  & $0$ \\
$\langle B' |D^{(8)}_{88}D^{(8)}_{33} | B \rangle$
  & $-\frac{\sqrt{6}}{40}$ & $0$
  & $\frac{\sqrt{3}}{20}$ & $0$  
  & $-\frac{\sqrt{2}}{45} T_{3}$ & $-\frac{1}{45\sqrt{2}} T_{3}$ 
  & $0$ \\
$\langle B' |d_{bc3} D^{(8)}_{8c}D^{(8)}_{3b}| B \rangle$
  & $-\frac{1}{10\sqrt{2}}$ & $\frac{1}{10\sqrt{2}} T_{3}$
  & $\frac{1}{10}$ & $-\frac{1}{10} T_{3}$
  & $-\frac{1}{9\sqrt{6}} T_{3}$ & $-\frac{7}{45\sqrt{6}} T_{3}$ 
  & $0$ \\
 \hline
 \hline
\end{tabular}
\end{center}
\end{table}
\begin{table}[htp]
\setlength{\tabcolsep}{5pt}
\renewcommand{\arraystretch}{2.2}
  \caption{The transition matrix elements of the single Wigner $D$ function operators
  coming from the $\overline{15}$-plet component of the baryon wavefunctions when $a=3$.}
  \label{tab:6}
\begin{center}
\begin{tabular}{ c | c c c c c c c}
 \hline 
  \hline 
$B \rightarrow B'$
  & $\Sigma_{c}^{+} \rightarrow \Lambda_{c}^{+}$ & $\Xi_{c}^{\prime}  \rightarrow \Xi_{c}$ &
  $\Sigma_{c}^{*+} \rightarrow \Lambda_{c}^{+}$ & $\Xi_{c}^{*}  \rightarrow \Xi_{c}$ &
  $\Sigma_{c}^{*} \rightarrow \Sigma_{c}$ & $\Xi_{c}^{*} \rightarrow \Xi_{c}^{\prime}$ &
  $\Omega_{c}^{*0} \rightarrow \Omega_{c}^{0}$\\
 \hline
$\langle B'_{\bm{\overline{15}}} |D^{(8)}_{33} | B \rangle$
  & $-\frac{1}{6\sqrt{30}}$ & $\frac{\sqrt{10}}{36} T_{3}$
  & $\frac{1}{6\sqrt{15}}$ & $-\frac{\sqrt{5}}{18} T_{3}$ 
  & $-\frac{1}{9\sqrt{5}} T_{3}$ & $-\frac{5}{9\sqrt{30}} T_{3}$
  & $0$ \\
$\langle B'_{\bm{\overline{15}}} |D^{(8)}_{38}J_{3} | B \rangle$  
  & $0$ & $0$ & $0$ & $0$ 
  & $-\frac{1}{3\sqrt{15}} T_{3}$ & $-\frac{5}{9\sqrt{10}} T_{3}$
  & $0$ \\
$\langle B'_{\bm{\overline{15}}} |d_{ab3}D^{(8)}_{3a}J_{b} | B \rangle$ 
  & $-\frac{1}{4\sqrt{30}}$ & $\frac{\sqrt{10}}{24} T_{3}$
  & $\frac{1}{4\sqrt{15}}$ & $-\frac{\sqrt{5}}{12} T_{3}$ 
  & $-\frac{1}{18\sqrt{5}} T_{3}$ & $-\frac{5}{18\sqrt{30}} T_{3}$
  & $0$ \\
$\langle B' |D^{(8)}_{33} | B_{\bm{\overline{15}}} \rangle$
  & $-\frac{1}{2\sqrt{15}}$ & $\frac{1}{6\sqrt{10}} T_{3}$
  & $\frac{1}{\sqrt{30}}$ & $-\frac{1}{6\sqrt{5}} T_{3}$ 
  & $-\frac{1}{9\sqrt{5}} T_{3}$ & $-\frac{5}{9\sqrt{30}} T_{3}$
  & $0$ \\
$\langle B' |D^{(8)}_{38}J_{3} | B_{\bm{\overline{15}}} \rangle$  
  & $0$ & $0$ & $0$ & $0$ 
  & $-\frac{1}{3\sqrt{15}} T_{3}$ & $-\frac{5}{9\sqrt{10}} T_{3}$
  & $0$ \\
$\langle B' |d_{ab3}D^{(8)}_{3a}J_{b} | B_{\bm{\overline{15}}} \rangle$ 
  & $-\frac{1}{4\sqrt{15}}$ & $\frac{1}{12\sqrt{10}} T_{3}$
  & $\frac{1}{2\sqrt{30}}$ & $-\frac{1}{12\sqrt{5}} T_{3}$ 
  & $-\frac{1}{18\sqrt{5}} T_{3}$ & $-\frac{5}{18\sqrt{30}} T_{3}$
  & $0$ \\
 \hline 
 \hline
\end{tabular}
\end{center}
\end{table}
\begin{table}[htp]
\setlength{\tabcolsep}{5pt}
\renewcommand{\arraystretch}{2.2}
  \caption{The transition matrix elements of the single Wigner $D$ function operators
  coming from the $\overline{24}$-plet component of the baryon wavefunctions when $a=3$.}
  \label{tab:7}
\begin{center}
\begin{tabular}{ c | c c c}
 \hline 
  \hline 
$B \rightarrow B'$
  & $\Sigma_{c}^{*} \rightarrow \Sigma_{c}$ & $\Xi_{c}^{*} \rightarrow \Xi_{c}^{\prime}$ 
  & $\Omega_{c}^{*0} \rightarrow \Omega_{c}^{0}$\\
 \hline
$\langle B'_{\bm{\overline{24}}} |D^{(8)}_{33} | B \rangle$
  & $-\frac{1}{90\sqrt{2}} T_{3}$ & $-\frac{1}{45\sqrt{3}} T_{3}$
  & $0$ \\
$\langle B'_{\bm{\overline{24}}} |D^{(8)}_{38}J_{3} | B \rangle$  
  & $\frac{1}{15\sqrt{6}} T_{3}$ & $\frac{2}{45} T_{3}$
  & $0$ \\
$\langle B'_{\bm{\overline{24}}} |d_{ab3}D^{(8)}_{3a}J_{b} | B \rangle$ 
  & $-\frac{1}{45\sqrt{2}} T_{3}$ & $-\frac{2}{45\sqrt{3}} T_{3}$
  & $0$ \\
$\langle B' |D^{(8)}_{33} | B_{\bm{\overline{24}}} \rangle$
  & $-\frac{1}{90\sqrt{2}} T_{3}$ & $-\frac{1}{45\sqrt{3}} T_{3}$
  & $0$ \\
$\langle B' |D^{(8)}_{38}J_{3} | B_{\bm{\overline{24}}} \rangle$  
  & $\frac{1}{15\sqrt{6}} T_{3}$ & $\frac{2}{45} T_{3}$
  & $0$ \\
$\langle B' |d_{ab3}D^{(8)}_{3a}J_{b} | B_{\bm{\overline{24}}} \rangle$ 
  & $-\frac{1}{45\sqrt{2}} T_{3}$ & $-\frac{2}{45\sqrt{3}} T_{3}$
  & $0$ \\
 \hline 
 \hline
\end{tabular}
\end{center}
\end{table}
\begin{table}[htp]
\setlength{\tabcolsep}{5pt}
\renewcommand{\arraystretch}{2.2}
  \caption{The matrix elements of the single and double 
  Wigner $D$ function operators when $a=4+i5$.}
  \label{tab:8}
\begin{center}
\begin{tabular}{ c | c c}
 \hline 
  \hline 
$B \rightarrow B'$ 
  & $\Sigma_{c}^{++} \rightarrow \Xi_{c}^{+}$ 
  & $\Sigma_{c}^{*++} \rightarrow \Xi_{c}^{+}$  \\
 \hline
$\langle B_{\bm{\overline{3}}} |D^{(8)}_{p3} | B_{\bm{6}}\rangle$
  & $\frac{1}{2\sqrt{6}}$ 
  & $-\frac{1}{2\sqrt{3}}$  \\
$\langle B_{\bm{\overline{3}}} |D^{(8)}_{p8} \hat{J}_{3} | B_{\bm{6}}\rangle$
  & $0$  & $0$ \\
$\langle B_{\bm{\overline{3}}} |d_{bc3} D^{(8)}_{pb} \hat{J}_{c} | B_{\bm{6}}\rangle$
  & $-\frac{1}{4\sqrt{6}}$ 
  & $\frac{1}{4\sqrt{3}}$ \\
$\langle B_{\bm{\overline{3}}} |D^{(8)}_{83}D^{(8)}_{p8} | B_{\bm{6}}\rangle$
  & $-\frac{1}{20\sqrt{6}}$ 
  & $\frac{1}{20\sqrt{3}}$  \\
$\langle B_{\bm{\overline{3}}} |D^{(8)}_{88}D^{(8)}_{p3} | B_{\bm{6}}\rangle$
  & $-\frac{1}{20\sqrt{6}}$ 
  & $\frac{1}{20\sqrt{3}}$  \\
$\langle B_{\bm{\overline{3}}} |d_{bc3} D^{(8)}_{8c}D^{(8)}_{pb}| B_{\bm{10}} \rangle$
  & $-\frac{1}{20\sqrt{2}}$ 
  & $\frac{1}{20}$  \\
 \hline
 \hline
\end{tabular}
\end{center}
\end{table}
\begin{table}[htp]
\setlength{\tabcolsep}{5pt}
\renewcommand{\arraystretch}{2.2}
  \caption{The transition matrix elements of the single Wigner $D$ function operators
  coming from the $\overline{15}$-plet component of the baryon wavefunctions when $a=4+i5$.}
  \label{tab:9}
\begin{center}
\begin{tabular}{ c | c c c c c c c}
 \hline 
  \hline 
$B \rightarrow B'$ 
  & $\Sigma_{c}^{++} \rightarrow \Xi_{c}^{+}$ 
  & $\Sigma_{c}^{*++} \rightarrow \Xi_{c}^{+}$ \\
\hline
$\langle B'_{\bm{\overline{15}}} |D^{(8)}_{p3} | B \rangle$
  & $\frac{1}{18\sqrt{10}}$ 
  & $-\frac{1}{18\sqrt{5}}$ \\
$\langle B'_{\bm{\overline{15}}} |D^{(8)}_{p8}J_{3} | B \rangle$  
  & $0$  & $0$ \\
$\langle B'_{\bm{\overline{15}}} |d_{ab3}D^{(8)}_{pa}J_{b} | B \rangle$
  & $\frac{1}{12\sqrt{10}}$ 
  & $-\frac{1}{12\sqrt{5}}$ \\
$\langle B' |D^{(8)}_{p3} | B_{\bm{\overline{15}}} \rangle$
  & $-\frac{1}{2\sqrt{15}}$ 
  & $\frac{1}{\sqrt{30}}$ \\
$\langle B' |D^{(8)}_{p8}J_{3} | B_{\bm{\overline{15}}}\rangle$
  & $0$  & $0$  \\
$\langle B' |d_{ab3}D^{(8)}_{pa}J_{b} | B_{\bm{\overline{15}}}\rangle$
  & $-\frac{1}{4\sqrt{15}}$ 
  & $\frac{1}{2\sqrt{30}}$  \\
 \hline 
 \hline
\end{tabular}
\end{center}
\end{table}
\begin{table}[htp]
\setlength{\tabcolsep}{5pt}
\renewcommand{\arraystretch}{2.2}
\caption{The matrix elements of the single and double Wigner $D$
  function operators when $a=4-i5$.} 
  \label{tab:10}
\begin{center}
\begin{tabular}{ c | c c c c c c}
 \hline 
  \hline 
$B \rightarrow B'$
  & $\Xi_{c}^{\prime 0} \rightarrow \Lambda_{c}^{+}$ & $\Omega_{c}^{0}
  \rightarrow  \Xi_{c}^{+}$ 
  & $\Xi_{c}^{* 0} \rightarrow \Lambda_{c}^{+}$ & $\Omega_{c}^{*0} \rightarrow \Xi_{c}^{+}$
  & $\Xi_{c}^{* +} \rightarrow \Sigma_{c}^{++}$ & $\Omega_{c}^{*0} \rightarrow \Xi_{c}^{\prime +}$ \\
 \hline
$\langle B' |D^{(8)}_{\Xi^{-}3} | B \rangle$
  & $-\frac{1}{4\sqrt{3}}$ & $\frac{1}{2\sqrt{6}}$
  & $\frac{1}{2\sqrt{6}}$ & $-\frac{1}{2\sqrt{3}}$
  & $-\frac{1}{5\sqrt{2}}$ & $-\frac{1}{5\sqrt{2}}$ \\
$\langle B' |D^{(8)}_{\Xi^{-}8} \hat{J}_{3} | B \rangle$
  & $0$ & $0$ & $0$ & $0$
  & $\frac{1}{5\sqrt{6}}$ & $\frac{1}{5\sqrt{6}}$ \\
$\langle B' |d_{bc3} D^{(8)}_{\Xi^{-}b} \hat{J}_{c}| B \rangle$
  & $\frac{1}{8\sqrt{3}}$ & $-\frac{1}{4\sqrt{6}}$
  & $-\frac{1}{4\sqrt{6}}$ & $\frac{1}{4\sqrt{3}}$
  & $\frac{1}{10\sqrt{2}}$ & $\frac{1}{10\sqrt{2}}$ \\
$\langle B' |D^{(8)}_{83}D^{(8)}_{\Xi^{-}8}| B \rangle$
  & $\frac{\sqrt{3}}{40}$ & $0$
  & $-\frac{\sqrt{6}}{40}$ & $0$
  & $\frac{1}{90\sqrt{2}}$ & $\frac{1}{30\sqrt{2}}$ \\
$\langle B' |D^{(8)}_{88}D^{(8)}_{\Xi^{-}3}| B \rangle$
  & $-\frac{1}{40\sqrt{3}}$ & $-\frac{1}{10\sqrt{6}}$
  & $\frac{1}{20\sqrt{6}}$ & $\frac{1}{10\sqrt{3}}$
  & $\frac{1}{90\sqrt{2}}$ & $\frac{1}{30\sqrt{2}}$ \\
$\langle B' |d_{bc3} D^{(8)}_{8c}D^{(8)}_{\Xi^{-}b}| B \rangle$
  & $\frac{1}{40}$ & $-\frac{1}{20\sqrt{2}}$
  & $-\frac{1}{20\sqrt{2}}$ & $\frac{1}{20}$
  & $\frac{7}{90\sqrt{6}}$ & $\frac{1}{30\sqrt{6}}$ \\
 \hline 
 \hline
\end{tabular}
\end{center}
\end{table}
\begin{table}[htp]
\setlength{\tabcolsep}{5pt}
\renewcommand{\arraystretch}{2.2}
  \caption{The transition matrix elements of the single Wigner $D$ function operators
  coming from the $\overline{15}$-plet component of the baryon wavefunctions when $a=4-i5$.}
  \label{tab:11}
\begin{center}
\begin{tabular}{ c | c c c c c c}
 \hline 
  \hline 
$B \rightarrow B'$
  & $\Xi_{c}^{\prime 0} \rightarrow \Lambda_{c}^{+}$ & $\Omega_{c}^{0} \rightarrow \Xi_{c}^{+}$
  & $\Xi_{c}^{* 0} \rightarrow \Lambda_{c}^{+}$ & $\Omega_{c}^{*0} \rightarrow \Xi_{c}^{+}$
  & $\Xi_{c}^{* +} \rightarrow \Sigma_{c}^{++}$ & $\Omega_{c}^{*0} \rightarrow \Xi_{c}^{\prime +}$ \\
\hline
$\langle B'_{\bm{\overline{15}}} |D^{(8)}_{\Xi^{-}3} | B \rangle$
  & $\frac{1}{4\sqrt{15}}$ & $-\frac{1}{6\sqrt{10}}$
  & $-\frac{1}{2\sqrt{30}}$ & $\frac{1}{6\sqrt{5}}$
  & $\frac{1}{9\sqrt{5}}$ & $\frac{1}{3\sqrt{30}}$ \\
$\langle B'_{\bm{\overline{15}}} |D^{(8)}_{\Xi^{-}8}J_{3} | B \rangle$  
  & $0$ & $0$ & $0$ & $0$ 
  & $\frac{1}{3\sqrt{15}}$ & $\frac{1}{3\sqrt{10}}$\\
$\langle B'_{\bm{\overline{15}}} |d_{ab3}D^{(8)}_{\Xi^{-}a}J_{b} | B \rangle$
  & $\frac{3}{8\sqrt{15}}$ & $-\frac{1}{4\sqrt{10}}$
  & $-\frac{3}{4\sqrt{30}}$ & $\frac{1}{4\sqrt{5}}$ 
  & $\frac{1}{18\sqrt{5}}$ & $\frac{1}{6\sqrt{30}}$\\
$\langle B' |D^{(8)}_{\Xi^{-}3} | B_{\bm{\overline{15}}} \rangle$
  & $-\frac{1}{4\sqrt{5}}$ & $0$
  & $\frac{1}{2\sqrt{10}}$ & $0$ 
  & $\frac{1}{9\sqrt{30}}$ & $0$\\
$\langle B' |D^{(8)}_{\Xi^{-}8}J_{3} | B_{\bm{\overline{15}}} \rangle$
  & $0$ & $0$ & $0$ & $0$ 
  & $\frac{1}{9\sqrt{10}}$ & $0$\\
$\langle B' |d_{ab3}D^{(8)}_{\Xi^{-}a}J_{b} | B_{\bm{\overline{15}}} \rangle$
  & $-\frac{1}{8\sqrt{5}}$ & $0$
  & $\frac{1}{4\sqrt{10}}$ & $0$ 
  & $\frac{1}{18\sqrt{30}}$ & $0$\\
 \hline
 \hline
\end{tabular}
\end{center}
\end{table}
\begin{table}[htp]
\setlength{\tabcolsep}{5pt}
\renewcommand{\arraystretch}{2.2}
  \caption{The transition matrix elements of the single Wigner $D$
    function operators coming from the $\overline{24}$-plet component
    of the baryon wavefunctions when $a=4-i5$.} 
  \label{tab:12}
\begin{center}
\begin{tabular}{ c | c c}
 \hline 
  \hline 
$B \rightarrow B'$ 
  & $\Xi_{c}^{* +} \rightarrow \Sigma_{c}^{++}$ & $\Omega_{c}^{*0} \rightarrow \Xi_{c}^{\prime +}$ \\
\hline
$\langle B'_{\bm{\overline{24}}} |D^{(8)}_{\Xi^{-}3} | B \rangle$
  & $\frac{\sqrt{2}}{45}$ & $\frac{1}{30\sqrt{3}}$ \\
$\langle B'_{\bm{\overline{24}}} |D^{(8)}_{\Xi^{-}8}J_{3} | B \rangle$  
  & $-\frac{2\sqrt{2}}{15\sqrt{3}}$ & $-\frac{1}{15}$ \\
$\langle B'_{\bm{\overline{24}}} |d_{ab3}D^{(8)}_{\Xi^{-}a}J_{b} | B \rangle$
  & $\frac{2\sqrt{2}}{45}$ & $\frac{1}{15\sqrt{3}}$ \\
$\langle B' |D^{(8)}_{\Xi^{-}3} | B_{\bm{\overline{24}}} \rangle$
  & $-\frac{1}{45\sqrt{3}}$ & $-\frac{1}{30\sqrt{3}}$ \\
$\langle B' |D^{(8)}_{\Xi^{-}8}J_{3} | B_{\bm{\overline{24}}}\rangle$
  & $\frac{2}{45}$ & $\frac{1}{15}$ \\
$\langle B' |d_{ab3}D^{(8)}_{\Xi^{-}a}J_{b} | B_{\bm{\overline{24}}}\rangle$
  & $-\frac{2}{45\sqrt{3}}$ & $-\frac{1}{15\sqrt{3}}$ \\
 \hline 
 \hline
\end{tabular}
\end{center}
\end{table}



\begin{thebibliography}{99}
\bibitem{PDG}
P.~A.~Zyla \textit{et al.} [Particle Data Group],
``Review of Particle Physics,''
PTEP \textbf{2020}, no.8, 083C01 (2020).


\bibitem{Can:2013tna} 
  K.~U.~Can, G.~Erkol, B.~Isildak, M.~Oka, and T.~T.~Takahashi,
  JHEP \textbf{05}, 125 (2014).

\bibitem{Bahtiyar:2015sga}
H.~Bahtiyar, K.~U.~Can, G.~Erkol and M.~Oka,
Phys. Lett. B \textbf{747}, 281-286 (2015).

\bibitem{Bahtiyar:2016dom} 
  H.~Bahtiyar, K.~U.~Can, G.~Erkol, M.~Oka, and T.~T.~Takahashi,
  Phys. Lett. B \textbf{772}, 121-126 (2017).

\bibitem{Bahtiyar:2019ykq} 
  H.~Bahtiyar, K.~U.~Can, G.~Erkol, M.~Oka and T.~T.~Takahashi, 
  JPS Conf. Proc. \textbf{26}, 022027 (2019).

\bibitem{Aaij:2012da} 
  R.~Aaij {\it et al.} [LHCb Collaboration],
  Phys. Rev. Lett. \textbf{109}, 172003 (2012).

\bibitem{Aaij:2013qja} 
  R.~Aaij {\it et al.} [LHCb Collaboration],
  Phys. Rev. Lett. \textbf{110}, no.18, 182001 (2013).

\bibitem{Aaij:2014esa} 
  R.~Aaij {\it et al.} [LHCb Collaboration],
  Phys. Rev. Lett. \textbf{113}, 032001 (2014).

\bibitem{Aaij:2014lxa} 
  R.~Aaij {\it et al.} [LHCb Collaboration],
  Phys. Rev. Lett. \textbf{113}, no.24, 242002 (2014).

\bibitem{Aaij:2014yka} 
  R.~Aaij {\it et al.} [LHCb Collaboration],
  Phys. Rev. Lett. \textbf{114}, 062004 (2015).

\bibitem{Aaij:2017nav} 
  R.~Aaij {\it et al.} [LHCb Collaboration],
  Phys. Rev. Lett. \textbf{118}, no.18, 182001 (2017).

\bibitem{LHCb:2020gge}
R.~Aaij \textit{et al.} [LHCb],
Phys. Rev. D \textbf{102}, no.7, 071101 (2020).

\bibitem{LHCb:2021ptx}
R.~Aaij \textit{et al.} [LHCb],
Phys. Rev. D \textbf{104}, no.9, 9 (2021).

\bibitem{Isgur:1989vq} 
  N.~Isgur and M.~B.~Wise,
  Phys. Lett. B \textbf{232}, 113-117 (1989).

\bibitem{Isgur:1991wq} 
  N.~Isgur and M.~B.~Wise,
  Phys. Rev. Lett. \textbf{66}, 1130-1133 (1991).

\bibitem{Georgi:1990um} 
  H.~Georgi,
  Phys. Lett. B \textbf{240}, 447-450 (1990).


\bibitem{Yang:2016qdz} 
  Gh.-S.~Yang, H.-Ch.~Kim, M.~V.~Polyakov and M.~Prasza{\l}owicz,
  ``Pion mean fields and heavy baryons,''
  Phys. Rev. D \textbf{94}, 071502 (2016).

\bibitem{Diakonov:2010tf} 
  D.~Diakonov,
  [arXiv:1003.2157 [hep-ph]].

\bibitem{Witten:1979kh} 
  E.~Witten,
  Nucl. Phys. B \textbf{160}, 57-115 (1979).

\bibitem{Witten:1983tx}
E.~Witten,
Nucl. Phys. B \textbf{223}, 433-444 (1983).

\bibitem{Pauli:1942kwa} 
W.~Pauli and S.~M.~Dancoff,
Phys. Rev. \textbf{62}, no.3-4, 85 (1942).

\bibitem{Skyrme:1961vq} 
  T.~H.~R.~Skyrme,
  Proc. Roy. Soc. Lond. A \textbf{260}, 127-138 (1961).

\bibitem{Diakonov:1987ty} 
D.~Diakonov, V.~Y.~Petrov and P.~V.~Pobylitsa,
Nucl. Phys. B \textbf{306}, 809 (1988).

\bibitem{Wakamatsu:1990ud}
M.~Wakamatsu and H.~Yoshiki,
Nucl. Phys. A \textbf{524}, 561-600 (1991).

\bibitem{Diakonov:1997sj}
D.~Diakonov,
[arXiv:hep-ph/9802298 [hep-ph]].

\bibitem{Kim:2017jpx} 
  H.-Ch.~Kim, M.~V.~Polyakov and M.~Praszalowicz,
  Phys. Rev. D \textbf{96}, no.1, 014009 (2017).

\bibitem{Kim:2017khv}
H.-Ch.~Kim, M.~V.~Polyakov, M.~Praszalowicz and G.~S.~Yang,
Phys. Rev. D \textbf{96}, no.9, 094021 (2017)
[erratum: Phys. Rev. D \textbf{97}, no.3, 039901 (2018)].

\bibitem{Kim:2018xlc}
J.~Y.~Kim, H.-Ch.~Kim and G.~S.~Yang,
Phys. Rev. D \textbf{98}, no.5, 054004 (2018).

\bibitem{Kim:2019rcx}
J.~Y.~Kim and H.-Ch.~Kim,
PTEP \textbf{2020}, no.4, 043D03 (2020).

\bibitem{Yang:2020klp}
G.~S.~Yang and H.-Ch.~Kim,
Phys. Lett. B \textbf{808}, 135619 (2020).

\bibitem{Yang:2018uoj}
G.~S.~Yang and H.-Ch.~Kim,
Phys. Lett. B \textbf{781}, 601-606 (2018).

\bibitem{Yang:2019tst}
G.~S.~Yang and H.-Ch.~Kim,
Phys. Lett. B \textbf{801}, 135142 (2020).

\bibitem{Kim:2018nqf}
J.~Y.~Kim and H.-Ch.~Kim,
Phys. Rev. D \textbf{97}, no.11, 114009 (2018).

\bibitem{Kim:2020uqo}
J.~Y.~Kim and H.-Ch.~Kim,
PTEP \textbf{2021}, no.2, 023D02 (2021).

\bibitem{Kim:2019wbg}
J.~Y.~Kim and H.-Ch.~Kim,
PTEP \textbf{2021}, no.6, 063D03 (2021).

\bibitem{Kim:2021xpp}
J.~Y.~Kim, H.-Ch.~Kim, G.~S.~Yang and M.~Oka,
Phys. Rev. D \textbf{103}, no.7, 074025 (2021).

\bibitem{Kim:2020nug}
J.~Y.~Kim, H.-Ch.~Kim, M.~V.~Polyakov and H.~D.~Son,
Phys. Rev. D \textbf{103}, no.1, 014015 (2021).

\bibitem{Yan:1992gz}
T.~M.~Yan, H.~Y.~Cheng, C.~Y.~Cheung, G.~L.~Lin, Y.~C.~Lin and H.~L.~Yu,
Phys. Rev. D \textbf{46}, 1148-1164 (1992)
[erratum: Phys. Rev. D \textbf{55}, 5851 (1997)].

\bibitem{Huang:1995ke}
M.~Q.~Huang, Y.~B.~Dai and C.~S.~Huang,
Phys. Rev. D \textbf{52}, 3986-3992 (1995)
[erratum: Phys. Rev. D \textbf{55}, 7317 (1997)].

\bibitem{Pirjol:1997nh}
D.~Pirjol and T.~M.~Yan,
Phys. Rev. D \textbf{56}, 5483-5510 (1997).

\bibitem{Cheng:2015naa}
H.~Y.~Cheng and C.~K.~Chua,
Phys. Rev. D \textbf{92}, no.7, 074014 (2015).

\bibitem{Rosner:1995yu}
J.~L.~Rosner,
Phys. Rev. D \textbf{52}, 6461-6465 (1995).

\bibitem{Tawfiq:1998nk}
S.~Tawfiq, P.~J.~O'Donnell and J.~G.~Korner,
Phys. Rev. D \textbf{58}, 054010 (1998).

\bibitem{Ivanov:1999bk}
M.~A.~Ivanov, J.~G.~Korner, V.~E.~Lyubovitskij and A.~G.~Rusetsky,
Phys. Rev. D \textbf{60}, 094002 (1999).

\bibitem{Albertus:2005zy}
C.~Albertus, E.~Hernandez, J.~Nieves and J.~M.~Verde-Velasco,
Phys. Rev. D \textbf{72}, 094022 (2005).

\bibitem{Nagahiro:2016nsx}
H.~Nagahiro, S.~Yasui, A.~Hosaka, M.~Oka and H.~Noumi,
Phys. Rev. D \textbf{95}, no.1, 014023 (2017).

\bibitem{Chen:2007xf}
C.~Chen, X.~L.~Chen, X.~Liu, W.~Z.~Deng and S.~L.~Zhu,
Phys. Rev. D \textbf{95}, no.1, 014023 (2017).

\bibitem{Azizi:2008ui}
K.~Azizi, M.~Bayar and A.~Ozpineci,
Phys. Rev. D \textbf{79}, 056002 (2009).

\bibitem{Can:2016ksz}
K.~U.~Can, G.~Erkol, M.~Oka and T.~T.~Takahashi,
Phys. Lett. B \textbf{768}, 309-316 (2017).

\bibitem{Christov:1995vm}
C.~V.~Christov, A.~Blotz, H.-Ch.~Kim, P.~Pobylitsa, T.~Watabe,
T.~Meissner, E.~Ruiz Arriola and K.~Goeke, 
Prog. Part. Nucl. Phys. \textbf{37}, 91-191 (1996).

\bibitem{Suh:2022guw}
J.~M.~Suh, Y.~S.~Jun and H.-Ch.~Kim,
[arXiv:2202.09066 [hep-ph]].

\bibitem{Adler:1968tw}
S.~L.~Adler,
Annals Phys. \textbf{50}, 189-311 (1968).

\bibitem{Rarita:1941mf}
W.~Rarita and J.~Schwinger,
Phys. Rev. \textbf{60}, 61 (1941).

\bibitem{Kim:2018cxv}
H.-Ch.~Kim,
J. Korean Phys. Soc. \textbf{73}, no.2, 165-178 (2018).

\bibitem{Belle:2021qip}
J.~Yelton \textit{et al.} [Belle],
Phys. Rev. D \textbf{104}, no.5, 052003 (2021).

\bibitem{Cheng:2006dk}
H.~Y.~Cheng and C.~K.~Chua,
Phys. Rev. D \textbf{75}, 014006 (2007).


\bibitem{Blotz:1992pw}
A.~Blotz, D.~Diakonov, K.~Goeke, N.~W.~Park, V.~Petrov and P.~V.~Pobylitsa,
Nucl. Phys. A \textbf{555}, 765-792 (1993).

\bibitem{Kim:1995mr}
H.-Ch.~Kim, A.~Blotz, M.~V.~Polyakov and K.~Goeke,
Phys. Rev. D \textbf{53}, 4013-4029 (1996).

\bibitem{Kim:1997ip}
H.-Ch.~Kim, M.~Praszalowicz and K.~Goeke,
Phys. Rev. D \textbf{57}, 2859-2870 (1998).

\bibitem{Kim:1997ts}
H.-Ch.~Kim, M.~V.~Polyakov, M.~Praszalowicz and K.~Goeke,
Phys. Rev. D \textbf{57}, 299-307 (1998).

\bibitem{Ledwig:2008ku}
T.~Ledwig, A.~Silva, H.-Ch.~Kim and K.~Goeke,
JHEP \textbf{07}, 132 (2008).

\bibitem{Meissner:1986js}
U.~G.~Meissner, N.~Kaiser and W.~Weise,
Nucl. Phys. A \textbf{466}, 685-723 (1987).

\bibitem{Ledwig:2008es}
T.~Ledwig, A.~Silva and M.~Vanderhaeghen,
Phys. Rev. D \textbf{79}, 094025 (2009).

\bibitem{FOCUS:2001qdu}
J.~M.~Link \textit{et al.} [FOCUS],
Phys. Lett. B \textbf{525}, 205-210 (2002).

\bibitem{CLEO:2001zwr}
M.~Artuso \textit{et al.} [CLEO],
Phys. Rev. D \textbf{65}, 071101 (2002).

\bibitem{Belle:2013htj}
Y.~Kato \textit{et al.} [Belle],
Phys. Rev. D \textbf{89}, no.5, 052003 (2014).

\bibitem{Belle:2014fde}
S.~H.~Lee \textit{et al.} [Belle],
Phys. Rev. D \textbf{89}, no.9, 091102 (2014).


\end{thebibliography}
\end{document}